\documentclass[a4paper,fleqn]{cas-dc}

\usepackage[authoryear,longnamesfirst]{natbib}
\usepackage{url}
\usepackage{xcolor}

\usepackage{hyperref}
\usepackage{multirow}
\usepackage{algorithmic}
\usepackage{textcomp}
\usepackage{multirow,multicol, makecell, booktabs}

\usepackage{graphicx}
\usepackage{amsmath,amssymb}
\usepackage{booktabs, makecell, tabularx}
\usepackage{rotating}
\usepackage[export]{adjustbox}
\usepackage{booktabs}
\usepackage{subcaption}
\usepackage{graphicx}
\usepackage{pifont}
\newcommand{\cmark}{\textcolor{black!80!black}{\ding{51}}}
\newcommand{\xmark}{\textcolor{black}{\ding{55}}}
\definecolor{newcolor}{rgb}{.8,.349,.1}
\def\tsc#1{\csdef{#1}{\textsc{\lowercase{#1}}\xspace}}
\tsc{WGM}
\tsc{QE}

\begin{document}
\let\WriteBookmarks\relax
\def\floatpagepagefraction{1}
\def\textpagefraction{.001}

\shorttitle{D-TrAttUnet: Toward Hybrid CNN-Transformer Architecture for Generic and Subtle Segmentation in Medical Images}    

\shortauthors{Bougourzi Fares et al.}  

\title[mode = title]{D-TrAttUnet: Toward Hybrid CNN-Transformer Architecture for Generic and Subtle Segmentation in Medical Images}  
\author[1]{Fares Bougourzi}
\ead{fares.bougourzi@junia.com}
\author[3,4] {Fadi Dornaika}
\cormark[1]
\ead{fadi.dornaika@ehu.eus}
\author[2]{Cosimo Distante}
\ead{cosimo.distante@cnr.it}

\author[5]{Abdelmalik Taleb-Ahmed}
\ead{Abdelmalik.Taleb-Ahmed@uphf.fr}

\address[1]{Junia, UMR 8520, CNRS, Centrale Lille, Univerity of Polytechnique Hauts-de-France, 59000 Lille, France}
\address[2]{Institute of Applied Sciences and Intelligent Systems, National Research Council of Italy, 73100 Lecce, Italy}
\address[3]{University of the Basque Country UPV/EHU, San Sebastian, Spain}
\address[4]{IKERBASQUE, Basque Foundation for Science, Bilbao, Spain}
\address[5]{Universit{\'e} Polytechnique Hauts-de-France, Université de Lille, CNRS, Valenciennes, 59313, Hauts-de-France, France}

\tnotetext[1]{Corresponding Author} 

%


\begin{abstract}
Over the past two decades, machine analysis of medical imaging has advanced rapidly, opening up significant potential for several important medical applications. As complicated diseases increase and the number of cases rises, the role of machine-based imaging analysis has become indispensable. It serves as both a tool and an assistant to medical experts, providing valuable insights and guidance. A particularly challenging task in this area is lesion segmentation, a task that is challenging even for experienced radiologists. The complexity of this task highlights the urgent need for robust machine learning approaches to support medical staff.
In response, we present our novel solution: the D-TrAttUnet architecture. This framework is based on the observation that different diseases often target specific organs. Our architecture includes an encoder-decoder structure with a composite Transformer-CNN encoder and dual decoders. The encoder includes two paths: the Transformer path and the Encoders Fusion Module path. The Dual-Decoder configuration uses two identical decoders, each with attention gates. This allows the model to simultaneously segment lesions and organs and integrate their segmentation losses.

To validate our approach, we performed evaluations on the Covid-19 and Bone Metastasis segmentation tasks. We also investigated the adaptability of the model by testing it without the second decoder in the segmentation of glands and nuclei. The results confirmed the superiority of our approach, especially in Covid-19 infections and the segmentation of bone metastases. In addition, the hybrid encoder showed exceptional performance in the segmentation of glands and nuclei, solidifying its role in modern medical image analysis.
\end{abstract}



\maketitle

\section{Introduction}
\label{sec:introduction}

The twenty-first century has witnessed a rise in both infectious and non-infectious diseases which have led to an increased number of cases and fatalities. This highlights the urgency for comprehensive healthcare approaches and preventive measures to address the escalating health crisis \cite{hambleton2023rising, baker2022infectious}.  Over the past two decades, medical imaging has demonstrated its effectiveness in early diagnosis, treatment planning, monitoring, research, education, and healthcare collaboration. It has transformed medical practice by providing insight into the inner workings of the human body, enabling better patient outcomes, and expanding medical knowledge. However, the increasing number of cases shows that there is a high demand to explore  automatic and effective machine learning approaches to ease the burden on medical staff.

The emergence and advancement of cutting-edge computational techniques, particularly within the realm of deep learning, have demonstrated their efficiency when abundant and well-labeled data are available \cite{shamshad2023transformers, sirinukunwattana2016locality}.  However, obtaining a sufficient amount of data with high labeling quality is still a challenge in the medical field due to many limitations. For example, the process of labeling medical images for segmentation is still time and resource consuming and requires the expertise of radiologists and physicians for accurate annotation \cite{shamshad2023transformers, sirinukunwattana2016locality}. To deal with this limitation, many approaches have been investigated including, semi-supervised learning \cite{lei2022semi}, self-supervised \cite{chaitanya2020contrastive},and data augmentation \cite{garcea2023data}.  However, most of these approaches contain several training stages, there is difficulty in choosing the proxy task in self-supervising and the difficulty to select the appropriate data augmentation techniques from one task to another, which requires extensive experiments \cite{garcea2023data}. 
In addition to the aforementioned techniques, there have been many attempts to develop an efficient deep learning architecture for medical imaging segmentation with dealing with the shortcoming of learning from limited data and to have a good generalization ability in different medical imaging segmentation tasks  \cite{wang2022mixed, petit_u-net_2021, wang2022uctransnet}.

In addressing the complex challenge of diseases affecting one or multiple organs, current state-of-the-art approaches have primarily focused on segmenting infection regions without considering the specific organ of interest  \cite{fan_inf-net_2020, wang_noise-robust_2020, paluru_anam-net_2021, liu_covid-19_2021, zhao2021scoat, wang_focus_2022, zhao2021scoat, wang_focus_2022}. Recognizing the limitations of such approaches, we undertake an exploration to devise a more efficient strategy for designing a highly effective deep learning architecture. Our emphasis lies in developing a model that thoroughly explores the segmentation of the tissues within the organ of interest to accurately identify infection regions. Our approach aims to provide a more comprehensive and targeted solution to the intricate nature of organic diseases.

Since the development of advanced deep learning approaches, particularly Convolutional Neural Networks (CNNs), they have become the dominant methods for medical imaging segmentation, including architectures like U-Net, Attention U-Net (AttUnet), and U-Net++ (Unet++). In recent years, the tremendous success of Transformers in natural language processing (NLP) tasks has spurred extensive investigations into their application for medical imaging segmentation, demonstrating promising performances. This exploration has given rise to Transformer-based segmentation architectures like Swin-UNet, characterized by a U-Net structure with a "U" shape and fully multi-head self-attention blocks, excluding convolutional blocks \cite{shamshad2023transformers}. Conversely, several works have endeavored to propose hybrid architectures, leveraging both Transformer and convolutional blocks, such as TransBTS \cite{wang2021transbts} and UNETR \cite{hatamizadeh_unetr_2022}. These efforts can be classified into three categories: (i) utilizing only Transformer layers to reconstruct the encoder \cite{hatamizadeh_unetr_2022}, (ii) employing a CNN encoder to extract deep representations, followed by the application of Transformer layers to the embedded CNN features \cite{wang2021transbts, wang2022mixed}, and (iii) integrating Transformer blocks in the skip connection phase to transmit encoder features to the decoder layers \cite{petit_u-net_2021, wang2022uctransnet}. Despite the variety of approaches that combine CNN and Transformer blocks, there is still room for improvement. Our approach aims to effectively merge both CNN and Transformer blocks in the encoding phase to simultaneously extract local, global, and long-range dependencies features. To achieve this, we propose a Transformer Encoder path and combine it with the extracted CNN features at different levels within a proposed Encoders Fusion Module.

In this paper, a new Transformer-CNN based approach is proposed based on the observation that diseases usually affect one or more human body organs. The proposed solution aims to direct training to the object of interest (the organ) by using a second decoder for organ segmentation as a secondary task. On the other hand, a hybrid Transformer-CNN encoder is proposed to extract rich features in the encoding phase, which plays a crucial role in avoiding the shortcut of CNN and Transformer approaches.
In fact, segmenting medical images is challenging due to the inherent diversity of disease mechanisms, characteristics, and effects. Our proposed approach is evaluated using two tasks with Organ/Infection propriety which are: Bone Metastasis Segmentation (BM), and Covid-19 Infection Segmentation (in both binary and multi-class scenarios). For these two tasks, the hybrid CNN-Transformer encoder and dual-decoder are exploited. Furthermore, Gland and Nuclei Segmentation are evaluated to check the effectiveness of the proposed hybrid CNN-Transformer approach for tasks that have no Organ/Infection property.
In summary, the main contributions of this work are:

\begin{itemize}
  \item  We propose a novel hybrid CNN-Transformer architecture by integrating multi-level Transformer features into the encoder. This approach aims to capture higher-level local features while maintaining long-range dependencies from diverse input patches.

  \item  Our proposed decoder consists of dual identical decoders to simultaneously address lesion and organ segmentation. Leveraging Attention Gates, Residual Blocks, and Upsampling layers, inspired by the observation that many diseases affect one or multiple human body organs.

  \item We evaluate our approach on challenging segmentation tasks, including Bone Metastasis and Covid-19 infection in both binary and multiclass scenarios. Additionally, we assess the effectiveness of the hybrid encoder in Gland and Nuclei segmentation, providing a comprehensive understanding of the model's versatility.

  \item  Extensive comparisons with various segmentation methods, including CNN and Transformer-based approaches, demonstrate the superior performance of our D-TrAttUnet architecture in Bone Metastasis and Covid-19 segmentation tasks. The hybrid Encoder proves effective in Gland and Nuclei segmentation. The code for D-TrAttUnet will be made publicly available at \url{https://github.com/faresbougourzi/D-TrAttUnet}.
\end{itemize}

This paper is organised as follows: Section \ref{S:2} presents some related work on CNN-based and Transformer-based segmentation architectures. In section \ref{S:3}, we describe our proposed approach. Section \ref{S:4} consists of the description of the datasets used to evaluate the performance of our approach. Section \ref{S:5} presents and discusses the experiments and results. Section \ref{S:6} shows visualization comparison between the prediction results of our approach and the best comparison approaches. Finally, section \ref{S:7} concludes this paper.

\section{Related Work}
\label{S:2}
In this section, we briefly review related work in the field of medical image segmentation that includes both CNN (Convolutional Neural Network) and Transformer-based approaches, then we will describe the chosen tasks to evaluate the performance of the proposed approach, with highlighting the importance and the challenges of each task.

\subsection{CNN Segmentation Architectures}

Since the great success of the first deep  CNN architecture ``Alexnet'' \cite{krizhevsky_imagenet_2012} in ImageNet \cite{deng_imagenet_2009} challenge in 2012, CNNs have reached the state of the art performance in many computer vision and machine learning tasks \cite{bougourzi_deep_2022, bougourzi_cnn_2022}. Segmentation tasks have been influenced by the great success of the CNNs and therefore many CNN architectures have proved their ability to segment many complicated medical imaging tasks \cite{ronneberger_u-net_2015, zhou_evolutionary_2020, tomar_fanet_2022}.  Since  Unet architecture  \cite{ronneberger_u-net_2015} was proposed in 2017, great progress has been made and a lot of Unet variants have been proposed such as  Attenion Unet (Att-Unet) \cite{oktay_attention_2018}, Unet++ \cite{zhou_unet_2018}, ResUnet \cite{zhang_road_2018}.

Unet \cite{ronneberger_u-net_2015}  is a CNN architecture with Encoder-Decoder structure. Unet's encoder consists of  consecutive CNN layers. Each layer contains convolutional and mapooling layers. On the other hand, the decoder consists of  consecutive decovolutional layers. The encoder and decoder are connected by skip connections, where encoder feature maps are concatenated with the decoder features to maintain fine-grained details by passing them to the decoder. This forms the ``U-shape''. In  \cite{oktay_attention_2018}, O. Oktay et al. proposed Attention Gate (AT) to determine the salient regions by using the encoder and decoder feature maps simultaneously.

\subsection{Transformers in CV}
Transformers are capable of capturing long-range  dependencies between sequence elements. Therefore, Transformers are widely used in the Natural Language Processing (NLP) domain \cite{khan_transformers_2021}. Inspired by the great success in the NLP domain, transformers have also been extensively studied in the computer vision domain in the last two years \cite{khan_transformers_2021}. Transformers have shown promising results in many computer vision tasks and many transformer-based architectures have been proposed such as ViT \cite{dosovitskiy_image_2020}, Swin-Transformer \cite{liu_swin_2021}, and Deit \cite{pmlr-v139-touvron21a}.

Similarly, Transformers have got much interest in Medical imaging domain \cite{shamshad_transformers_2022}. Indeed, Transformers have shown promising performance in many medical imaging tasks such as classification \cite{dai_transmed_2021}, detection \cite{shen_cotr_2021} and segmentation \cite{hatamizadeh_unetr_2022}. Since the focus of this work is segmentation task,  some transformer-based segmentation approaches will be described. The segmentation architectures can be classified as 2-D \cite{wu_fat-net_2022, petit_u-net_2021} or 3-D modalities \cite{hatamizadeh_unetr_2022, wang_transbts_2021}.

In \cite{wu_fat-net_2022},  H. Wu et al. proposed a CNN-Transformer architecture called ``Fat-Net'', where two encoders (CNN and Transformer encoders) are used. The feature maps of the two encoders were concatenated to have richer features from the two encoders.  The Squeeze and Excitation (SE) module \cite{hu_squeeze-and-excitation_2018} was applied on the concatenated features to  identify the most important feature correlations from different feature channels. Fat-Net was evaluated for skin lesion segmentation using four public datasets.   In the U-Transformer \cite{petit_u-net_2021} architecture, Multi-Head Self-Attention and  Cross Attention modules were injected into the U-Net architecture. These two modules were placed at the skip connection to learn the global context information from the U-Net encoder and pass them to its decoder \cite{petit_u-net_2021}. The U-Transformer architecture performed well in two abdominal CT-image datasets \cite{petit_u-net_2021}. In  \cite{hatamizadeh_unetr_2022}, A. Hatamizadeh et al. proposed a transformer-based architecture for multi-classes 3D segmentation called ``UNETR''. The encoder of UNETR  was constructed by a transformer, from which  four levels features are obtained and rescaled by deconvolutional layers. The rescaled maps  were connected to the CNN decoder via skip connections at different resolutions, forming the ``U-shape''.

\textcolor{black}{Several approaches have been proposed to integrate CNN and Transformer blocks into single architectures, with a focus on incorporating Transformer architectures into the encoding block. The following encoder configurations have been suggested: constructing the encoder solely using Transformer architecture \cite{hatamizadeh_unetr_2022, zhu2023brain}; employing two parallel encoders, one based on Transformer and the other on CNN, and then combining their outputs \cite{wu_fat-net_2022, he2023medical}; or implementing a CNN encoder followed by Transformer blocks \cite{wang2021transbts, wang2022mixed}. Despite the variety of approaches that combine CNN and Transformer blocks, there is still room for improvement. Our approach aims to effectively merge both CNN and Transformer blocks in the encoding phase to simultaneously extract local, global, and long-range dependencies features. To achieve this, we propose a Transformer Encoder path and combine it with the extracted CNN features at different levels within a proposed Encoders Fusion block.}

\subsection{Introducing the Evaluated Medical Imaging Segmentation Tasks}

\textcolor{black}{BM segmentation is exceptionally demanding, even for seasoned radiologists, owing to several complexities. The omnipresence of bone throughout the body makes allocating and tracking BM lesions arduous and time-consuming \cite{heindel2014diagnostic, afnouch2023bm}. Furthermore, BM exhibits significant appearance variability, contingent on lesion nature, bone location, and infection progress and stage. Distinguishing BM from other benign conditions, fractures, bone islands, and degenerative changes is often confounding \cite{heindel2014diagnostic, afnouch2023bm}. 
On the other hand, automatic segmentation of Colon Glands from Histology Images is critical for cancer grading, a crucial step in determining cancer progression and the appropriate treatment plan to save lives \cite{sirinukunwattana2017gland}. Traditionally, cancer grading has relied on subjective and time-consuming assessments by pathologists that require manual quantification of tumor cell abnormalities. These challenges underscore the compelling need for automated methods that use machine learning approaches \cite{le2021joint}.}

\textcolor{black}{In addition to BM and Gland segmentation, our approach extends to Covid-19 infection segmentation. This task encompasses both binary segmentation, discerning infection presence or absence \cite{fan_inf-net_2020, wang_noise-robust_2020, paluru_anam-net_2021, liu_covid-19_2021}, and multi-class segmentation, which provides a nuanced view by categorizing non-infection, Ground-Glass Opacities (GGO), or Consolidation \cite{zhao2021scoat, wang_focus_2022}. Binary segmentation quantifies infection spread in the lungs, while multi-class segmentation offers insights into the infection's stage, progress, and severity \cite{zhao2021scoat, wang_focus_2022, bougourzi_ilc-unet_2022}. However, the scarcity of data for multi-class Covid-19 segmentation has limited research in this area \cite{zhao2021scoat, wang_focus_2022}. The main challenge in segmenting Covid-19 infections arises from their high variability in intensity, shape, position, and type, further complicated by factors such as infection stage, symptoms, and severity \cite{kumar_singh_lunginfseg_2021, Laradji_2021_WACV}. These challenges necessitate an efficient deep learning approach for effectively segmenting Covid-19 infection to save the patients live.}

In this section, we emphasize the tasks chosen to evaluate our proposed approach and compare it with other methods, known for their challenging nature. The primary objective of this study is to assess the generalization capabilities of various recently proposed approaches and compare their performance with our own. Specifically, our aim here is to study the generalization ability of our proposed approach across a variety of tasks and compare it with the behavior of state-of-the-art approaches. Our approach is specifically tailored to offer an efficient solution for medical imaging segmentation tasks.


\section{The Proposed Approach}
\label{S:3}

\begin{figure}[htp]
    \includegraphics[width =3.5in, height = 2.2in] {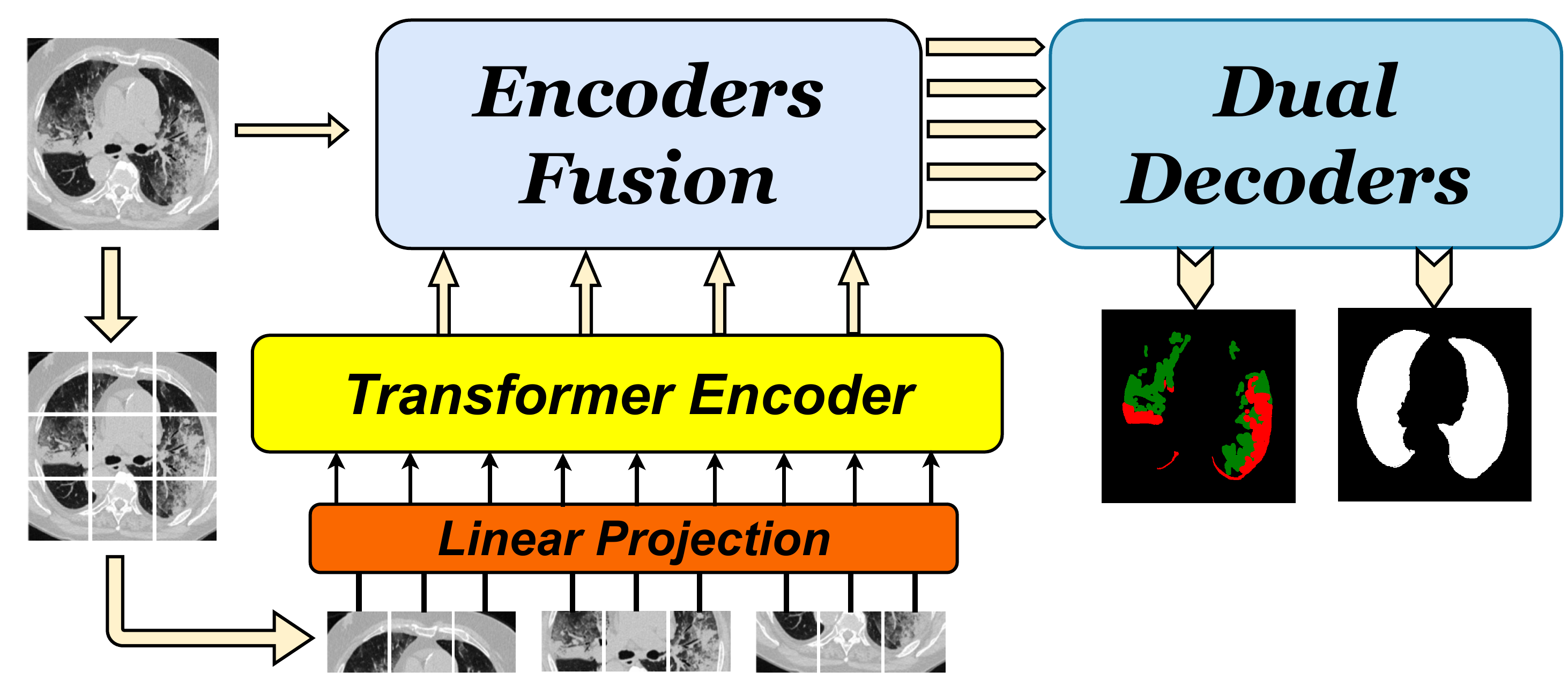} 
    \caption{The summary of our proposed D-TrAttUnet approach. }
    \label{fig:approachsum}
\end{figure}

\begin{figure*}
    \includegraphics[width =7in, height = 4.6in] {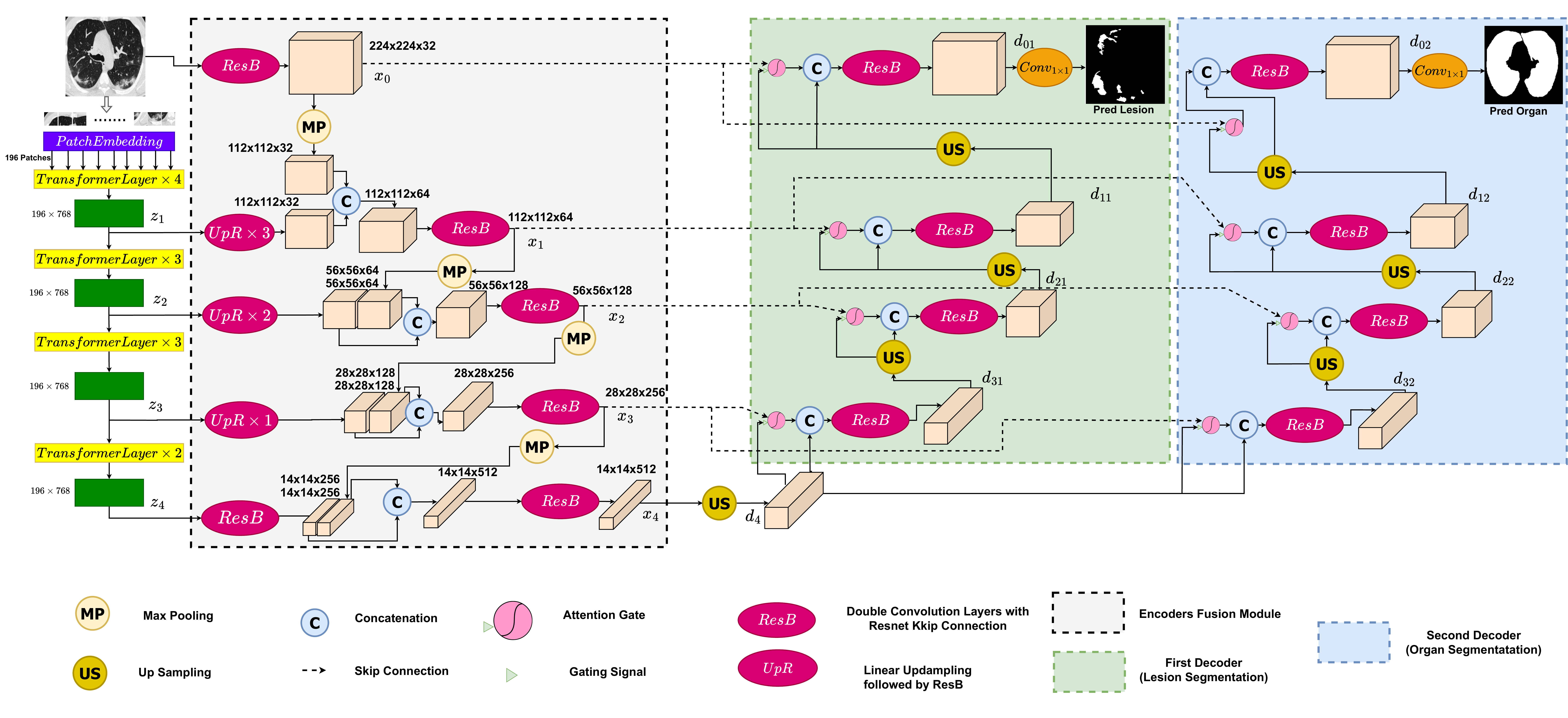} 
    \caption{Detailed Structure of the proposed D-TrAttUnet architecture. }
    \label{fig:approachA}
\end{figure*}

Our proposed approach for Lesion-Organ segmentation is presented in Figure \ref{fig:approachsum}. We introduce the D-TrAttUnet architecture, a novel compound CNN-Transformer architecture with a U-Net-like structure that leverages Attention Gates (AG). In Figure \ref{fig:approachA}, we provide a detailed illustration of the D-TrAttUnet's architecture. The key feature of our D-TrAttUnet lies in its encoder, which utilizes both transformer layers and ResBlocks (ResB) to extract rich and diverse features. Given the high variability in the shape, size, and position of medical imaging pathologies \cite{sun_systematic_2020, kumar_singh_lunginfseg_2021, Laradji_2021_WACV}, it is crucial to capture diverse features from the medical images. To achieve this, our approach combines locally extracted features using CNN filtering with globally aggregated features from the image patches through transformer layers. This allows the model to effectively handle the varying characteristics of medical pathologies and provide efficient segmentation.

In our proposed D-TrAttUnet, the encoder has two paths: the Unet-like path and the Transformer path. The input image $x \in \mathbb{R}^{H\times W\times C}$, where $H$, $W$ and $C$ are the height, the width and the input channels, is fed into both paths. 

\subsection{Transformer Encoder}

In the Transformer Encoder, the input tensor $x$ is divided into uniform, non-overlapping patches denoted as $x_v \in \mathbb{R}^{N \times (S^2 \times C)}$, where each patch has a size of $S \times S \times C$, and $N$ is the total number of patches given by $N = \frac{{H \times W}}{{S^2}}$. These patches undergo linear transformation denoted by $E \in \mathbb{R}^{(S^2 \times C) \times K}$ to project them into an embedding space $z_0$, where $K$ represents the dimensionality of the embedding space. The embedding $z_0$ is computed as follows:

\begin{equation}
\label{eq:1}
z_0 = [x_v^1 E;x_v^2 E;.........;x_v^N E]
\end{equation}


The embedded features $z_0 \in \mathbb{R}^{N \times K}$ are then fed into Transformer layers, similar to previous works \cite{dosovitskiy_image_2020, vaswani_attention_2017}. Each Transformer layer comprises two Layernorm (LN) blocks, a Multi-Head Self-Attention (MSA) block, a multi-layer perceptron (MLP) block, and residual connections. For the $l$-th Transformer layer, the embedded input features $z_{l-1}$ are processed as follows:

\begin{equation}
\label{eq:2}
z_l^\prime = MSA(\;LN(z_{l-1})\;) + z_{l-1}
\end{equation}

Here, $\text{LN}(z_{l-1})$ represents the output of the Layernorm block and $\text{MSA}$ denotes the Multi-Head Self-Attention block. The MSA operation is defined as:

\begin{equation}
\label{eq:3}
\text{MSA}(s) = ([SA_1(s), SA_2(s), \ldots , SA_h(s)]) \, U_{\text{msa}}
\end{equation}

where $SA_1, SA_2, \ldots , SA_h \in \mathbb{R}^{N \times \frac{K}{h}}$ are Self-Attention heads, and $U_{\text{msa}} \in \mathbb{R}^{K \times K}$ is the weighting matrix for the SA features.

The output $z_l^\prime$ is then further processed as follows:

\begin{equation}
\label{eq:4}
z_l = MLP(\;LN(z_l^\prime)\;) + z_l^\prime
\end{equation}

where $MLP$ consists of  two Linear layers with a GELU non-linearity. The first Linear layer ($MLP_1  \in \mathbb{R}^{ K\times K_{MLP}}$) projects $LN(z_l^\prime)$ into $K_{MLP}$, then the second  Linear layer ($MLP_2  \in \mathbb{R}^{ K_{MLP}\times K}$) projects the features into a $K$-dimensional space.

In our approach, we set $L = 12$, $h = 12$, $K = 768$, $K_{\text{MLP}} = 3072$, and $S = 16 \times 16$ pixels. Thus, for an image resolution of $W = H = 224$, the number of patches is $N = 196$.

\subsection{Encoders Fusion Module}

\textcolor{black}{To harness multi-scale features from different Transformer layers (stages), we meticulously selected embedded features from layers 4, 7, 10, and 12, denoted as $Tr_{1}$, $Tr_{2}$, $Tr_{3}$, and $Tr_{4}$, respectively. The decision to choose  Transformer features from different levels allows for the extraction of diverse and informative features, instead of using features from closely located layers, which might convey many similarities in features. By focusing on these chosen layers, our aim is to capture a spectrum of scales and complexities inherent in the input data. This strategy aims to enhance the network's capability to handle variations in the medical imaging context effectively. The selected layers yield embeddings with a consistent shape of $196\times 768$. Before being fed into the decoders, these features undergo filtering and fusion using CNN-based operations, as detailed below.}

\begin{figure*}[htp]
\centering
    \includegraphics[width = 6in, height = 3.8in] {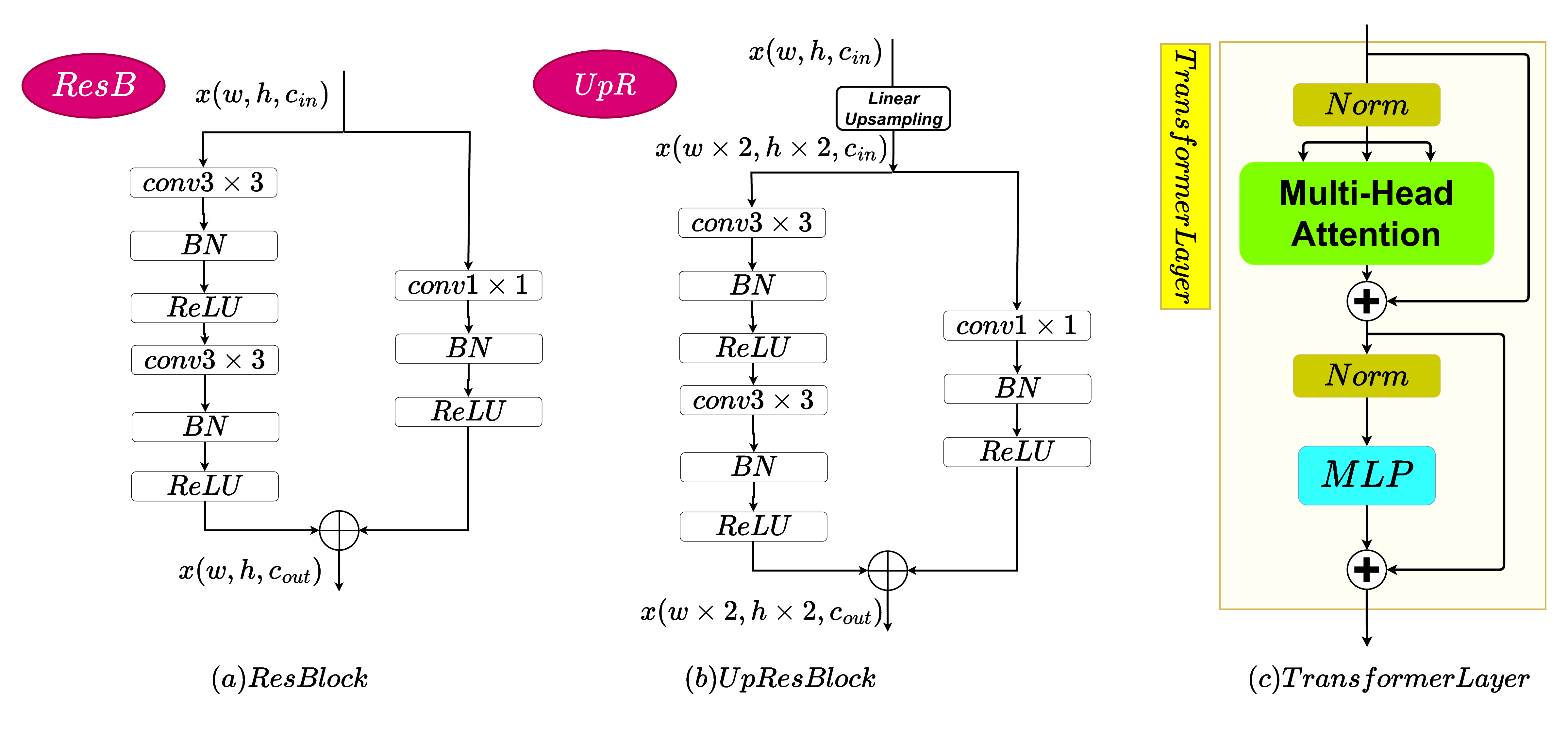} 
    \caption{Description of ResBlock (ResB), UpResBlock (UpR) and TransformerLayer. }
    \label{fig:blocks}
\end{figure*}

To obtain a 3D shape for the intermediate results of the transformer (a sequence of vectors), $z_l$ is reshaped to $ 14\times 14\times 768$, since $14\times 14= 196$. These reshaped features  for $Tr_{1}$,  $Tr_{2}$, $Tr_{3}$ and $Tr_{4}$ are denoted by $z_{1}$,  $z_{2}$, $z_{3}$ and $z_{4}$, respectively. To inject the transformer features into different layers of D-TrAttUnet and combine them with the CNN layers, UpResBlock (UpR) is introduced as depicted in Figure \ref{fig:blocks}-b.  UpR consists of linear upsampling followed by ResBlock (ResB)  as depicted in Figure  \ref{fig:blocks}-a. ResB consists of two 3 by 3 convolutional block, each followed by Batch Normalization and ReLU activation function. In addition,  the output of the two convolutional layers are summed with the input passed to residual connection, which consists of 1 by 1 convolutional block, followed by Batch Normalization and ReLU activation function, as shown in equations (\ref{eq:5}) and (\ref{eq:6}): 

\begin{equation}
\label{eq:5}
x_{out_1} = ReLU(\;BN(Conv3\times3_1(x_{in})\;) 
\end{equation}

\begin{equation}
\label{eq:6}
\begin{split}
x_{out} = ReLU(\;BN(Conv3\times3_2(x_{out_1}))) + \\ ReLU(\;BN(Conv1\times1(x_{in})))
\end{split}
\end{equation}

where $Conv3\times3_1 \in \mathbb{R}^{3\times 3\times C_{out}}$, $Conv3\times3_2 \in \mathbb{R}^{3\times 3\times C_{out}}$ and $Conv1\times1 \in \mathbb{R}^{1\times 1\times C_{out}}$.

As shown in Figure \ref{fig:approachA}, the Encoders Fusion Module has four layers. The first layer uses ResB on the input image $x \in \mathbb{R}^{H\times W\times C}$ to obtain the first feature maps as shown in equation (\ref{eq:11}).

\begin{equation}
\label{eq:11}
x_{0} = ResB (x)
\end{equation}

The Encoders Fusion Module consists of feature extraction and fusion processes at four layers, resulting in feature sets denoted as $x_1$, $x_2$, $x_3$, and $x_4$. These features are obtained through extraction and fusion of both CNN and Transformer features. To elaborate, the first Encoders Fusion features ($x_1$) are created by combining the CNN features ($x_0$) with the features from the first Transformer layer ($z_1$), as shown in Equation (\ref{eq:7}). This process entails passing $x_0$ through a max-pooling layer (MP) and sending $z_1$ through three consecutive UpR blocks ($UpR^3$) to capture higher-level features that align with the max-pooled $x_0$ features. The resulting CNN and Transformer features are concatenated and then processed through a ResB block, to extract richer features containing both features types.

\begin{equation}
\label{eq:7}
x_1 = ResB \; (Concatenate \;(UpR\;^3(z_{1}), MP(x_0))\;)
\end{equation}

Subsequently, the second Encoders Fusion features ($x_2$) are generated by fusing the previous Encoders Fusion features ($x_1$) with the features from the second Transformer layer ($z_2$), as detailed in Equation (\ref{eq:8}). This involves applying two UpR blocks to the Transformer features ($z_2$), max-pooling $x_1$, then concatenating the results before passing them through a ResB block.
Similarly, $x_3$ is obtained by fusing $x_2$ and the features from the third Transformer layer ($z_3$) using a combination of max-pooling, an UpR block, concatenation, and a ResB block, as illustrated in Equation (\ref{eq:9}).
Finally, the ultimate Encoders Fusion features ($x_4$) are formed by merging $x_3$ with the features from the fourth Transformer layer ($z_4$), involving two ResB blocks, max-pooling, and concatenation, as demonstrated in Equation (\ref{eq:10}).
In summary, this process outlines how each set of features ($x_1$, $x_2$, $x_3$, and $x_4$) is obtained through a sequence of fusion and transformation operations, leveraging both CNN and Transformer features at different stages.

\begin{equation}
\label{eq:8}
x_2 = ResB \; (Concatenate \;(UpR\;^2(z_{2}), MP(x_1))\;)
\end{equation}

\begin{equation}
\label{eq:9}
x_3 = ResB \; (Concatenate \;(UpR\;(z_{3}), MP(x_2))\;)
\end{equation}

\begin{equation}
\label{eq:10}
x_4 = ResB \; (Concatenate \;(ResB\;(z_{4}), MP(x_3))\;)
\end{equation}

\begin{figure*}[htp]
\centering

\includegraphics[width = 5in, height=1.3in] {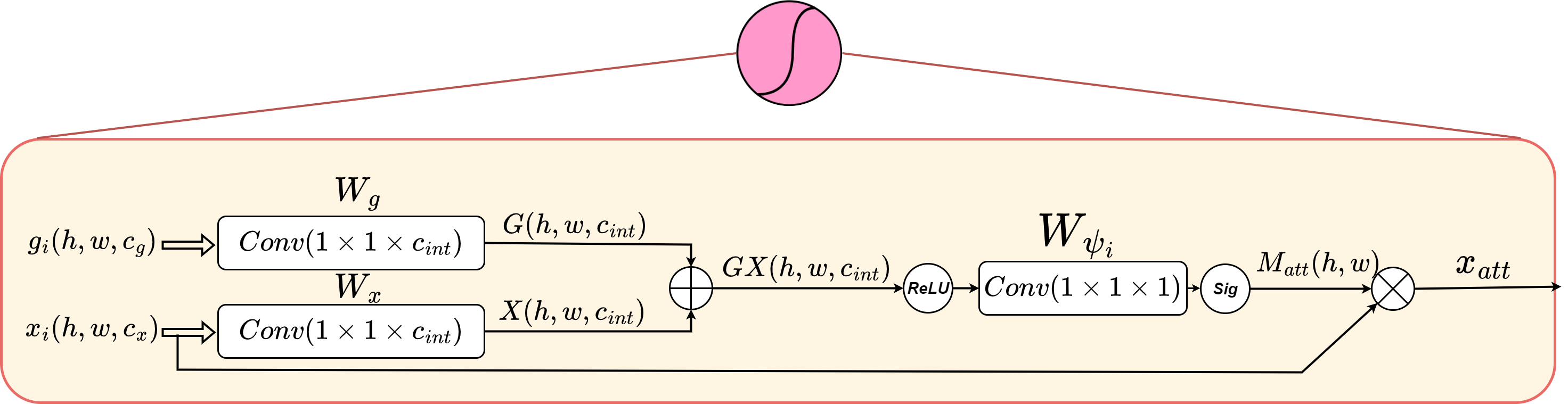}

\label{fig:attgate}
\caption{Attention Gate block, where $g_i$ is the gating signal and the $x_i$ is the input feature maps. $M_{att}(h,w)$ is the obtained  spatial attention, which is applied for all channels of the input feature maps ($x_i$). }
    \label{fig:att}
\end{figure*}

\subsection{Dual-Decoders}

In our proposed D-TrAttUnet architecture, we employ dual decoders. As illustrated in Figure \ref{fig:approachA}, the primary objective of the first decoder is to segment the lesion, whether it is Bone-Metastasis or Covid-19 infection. \textcolor{black}{Meanwhile, the second decoder focuses on segmenting the organ of interest, such as the Lung in the case of Covid-19 and the Bone in the context of Bone Metastasis (BM). The rationale behind incorporating a dedicated decoder for organ segmentation is twofold: firstly, this design facilitates the encoder's concentration within the organs, which are the primary locations of infections, where the two tasks losses  affect the common encoder. Secondly, it compels the model to distinguish between various tissues in CT-scans, a critical consideration given that tissues outside the organs may exhibit visual similarities to infection. This strategic inclusion of multi-tasking in the decoding phase, with a secondary task for organ segmentation, is intended to encourage the encoder to learn more diverse features, thereby enhancing the efficiency of the segmentation task.}

The bottleneck feature maps ($x_{5}$) of the  encoder are fed into the first expansion layer of the two decoders.
First, $x_{5}$ is Upsampled using a linear transformation to obtain  $d_{5}$, and then passed to the two decoders as shown in equation (\ref{eq:161}). On the other hand, the encoder feature maps $x_{1}, x_{2}, x_{3}$ and  $x_{4}$  are fed to the two decoders layers of D-TrAttUnet through skip connections, as shown in Figure \ref{fig:approachA}.  Following the Att-Unet architecture \cite{oktay_attention_2018}, three linear upsampling layers (US), four decoder layers, four attention gates, and four ResBlocks are used for each decoder, as shown in the following equations:

\begin{equation}
\label{eq:161}
d_{4} = US(x_{4})
\end{equation}

\begin{equation}
\label{eq:16}
d_{31} = ResB(\;Concatenate \; [AttGate(x_{3}, US(x_{4})), US(x_{4}]))
\end{equation}

\begin{equation}
\label{eq:17}
d_{21} = ResB(\;Concatenate \; [AttGate(x_{2}, US(d_{31})), US(d_{31}]))
\end{equation}

\begin{equation}
\label{eq:18}
d_{11} = ResB(\;Concatenate \; [AttGate(x_{1}, US(d_{21})), US(d_{21}]))
\end{equation}
\begin{equation}
\label{eq:19}
d_{01} = ResB(\;Concatenate \; [AttGate(x_{0}, US(d_{11})), US(d_{11}]))
\end{equation}

Similarly, $d_{32}, d_{22}, d_{12} $and $ d_{02}$ are obtained for the organ segmentation decoder. Finally, two convolutional 1 by 1 layers are used to match the feature map dimension of $d_{01}$ and $d_{02}$ to the lesion and organ masks prediction, which consist of a single channel for the organ and binary segmentation and three channels for the multi-classes Covid-19 segmentation.

The Attention Gate (AG) is depicted in Figure \ref{fig:att} \cite{oktay_attention_2018}, and it is defined as follows:

\begin{equation}
\label{eq:20}
M_{att} = \psi_{i} \; ( ReLU ( BN(\;W_{x} \; x_{i}) + BN(W_{g} \;g_{i})))
\end{equation}

where $W_{x} \in \mathbb{R}^{1\times 1\times C_{int}}$ and $W_{g} \in \mathbb{R}^{1\times 1\times C_{int}}$ are two linear transformations that transform the channels $c_{x}$ and $c_{g}$ from  $x_{i}$ and $g_{i}$, respectively, to $c_{int}$. $ \psi_{i} $ consists of $W_{\psi_{i}} \in \mathbb{R}^{1\times 1\times 1}$ followed by BatchNormalization (BN) and sigmoid activation function to learn the spatial attention coefficient $M_{att_{p}}$ for each pixel. The obtained spatial coefficients $M_{att}$ are applied to the skip feature maps of the encoder  $x_{i}$.

\begin{equation}
\label{eq:21}
x_{att} = M_{att} \; \otimes x_{i}
\end{equation}

Since our approach exploits multi-tasks approach with dual-decoders, the used loss function is defined by:


\begin{equation}
\label{eq:loss2}
\mathcal{L}_{Hybrid} = \alpha \, \mathcal{L}_{{LES-CE}}+\beta \, \mathcal{L}_{{LES-Dice}}+\gamma \,  \mathcal{L}_{O-CE}
\end{equation}

\textcolor{black}{
where $\mathcal{L}_{{LES-CE}}$ and $\mathcal{L}_{{LES-Dice}}$ are the CE and Dice losses of the lesion branch and $\mathcal{L}_{O-CE}$ is the loss associated with organ segmentation branch using CE loss function. 
The weights $\alpha$ and $\beta$  are set to 0.5 and 0.5, respectively, and $\gamma$ is set to 0.3 for the secondary task loss (organ segmentation).}

\section{Datasets}
\label{S:4}


Three tasks have been used to evaluate the performance of our approach. First, for BM segmentation, we utilized the BM-Seg dataset \cite{afnouch2023bm}. This dataset comprises data from 23 CT scans of 23 patients diagnosed with bone metastasis. A total of 1517 slices were confirmed to exhibit bone metastasis infections by three expert radiologists and were labeled for bone metastasis and bone regions masks. We followed the same splits as described in \cite{afnouch2023bm}, where five-fold cross-validation evaluation scenario were introduced.

For the Covid-19 infection segmentation task, we evaluated both binary and multi-class segmentation tasks, as summarized in Table 1. Dataset\_1 \cite{COVID-19-Dataset} consists of 100 slices showing Covid-19 infection, including lungs and multi-class infection masks (GGO and Consolidation). Dataset\_2 \cite{COVID-19-Dataset} comprises nine 3D CT scans, totaling 829 slices, with 373 slices indicating Covid-19 infection. Expert radiologists labeled this dataset, providing masks for lungs, binary infection (non-infected and infected), and multi-class labels (non-infected, GGO, and Consolidation).

For the binary segmentation task, we divided Dataset\_2 into 70\%-30\% splits for training and testing, respectively. For the multi-class segmentation task, we used Dataset\_2 and 50\% of Dataset\_1 (50 slices) for training, while the remaining 50 slices of Dataset\_1 were used for testing. Table 2 summarizes the number of slices for the GGO and Consolidation classes in both the training and testing splits. As shown in the table, the limited number of slices for each class presents a significant challenge for multi-class segmentation.

For Gland and Nuclear segmentation tasks, we employed the Gland segmentation dataset (GlaS) \cite{siri2017gland} and the MoNuSeg dataset \cite{kumar2019multi}, respectively. The Gland segmentation dataset (GlaS) has consists of 165 images, and the MoNuSeg dataset consists of 44 images. Following \cite{wang2022uctransnet}, three times five-fold cross-validation is performed for each task, and the mean and std results are considered.
\begin{table}
 \caption{The Summary of the used Covid-19 datasets. In which, we highlighted the number of CT-scans and slices for each dataset. }
\label{tab:datasets}
\centering
\begin{tabular}{|p{1.3cm}|p{4.2cm}|p{1.1cm}|p{0.9cm}|}
\hline

 \textbf{Name}  & \textbf{Dataset} &\textbf{\#CT-Scans} &   \textbf{\#Slices} \\\hline
Dataset\_1& COVID-19 CT segmentation \cite{COVID-19-Dataset}& 40& 100  \\ \hline 
Dataset\_2&  Segmentation dataset nr. 2 \cite{COVID-19-Dataset}&9 &  829  \\ \hline 

\end{tabular}
\end{table}

\begin{table*}
 \caption{\textcolor{black}{Data splitting of Covid-19 datasets for Multi-classes Covid-19 infection segmentation task. In which, the  number of slices (in brackets the number of CT-scans), infected slices among the total, slices with GGO infection, and slices with consolidation specified for each split (Train and Test).}  }
\label{tab:evalsce}
\centering
\scalebox{0.9}{\begin{tabular}{|p{1cm}|p{3cm}|p{1.8cm}|p{2cm}|p{2.5cm}|p{2.5cm}|}
\hline
\textbf{Split} &\textbf{Involved Dataset} & \textbf{Total \# Slices} &\textbf{\# Infected Slices} &   \textbf{\#Slices with GGO}& \textbf{\#Slices with Consolidation} \\\hline
 
Train& Dataset\_2 \newline 50\% of Dataset\_1 &879 (9+20 CT-scans) & 422& 345&272   \\ \hline
Test& 50\% of Dataset\_1 &50 (20 CT-scans)&50&  50& 40\\
\hline
\end{tabular}}
\end{table*}

\section{Experiments and results}
\label{S:5}
\subsection{Experimental~Setup}
To produce our experiments, we mainly used PyTorch \cite{paszke_pytorch_2019} library for deep learning. Each architecture is trained for 100 epochs with an initial learning rate of 0.1 and Adam optimizer. The batch size is set to 16 images. The used machine has NVIDIA RTX A5000 GPU with 24 GB of memory, 11th Gen Intel(R) Core(TM) i9-11900KF (3.50GHz) CPU and 64 of RAM. Three types of active data augmentation are used; random rotate with an angle between $-35^{\circ}$ and $35^{\circ}$ with a probability of 10\% and  random Horizontal  and vertical Flipping with probability of 20\% for each. 

\subsection{Evaluation Measurements}

\textcolor{black}{In order to compare the performance of our approach with the state-of-the-art approaches, the following evaluation metrics have been used: F1-score (F1-S), Dice score (D-S), Intersection Over Union (IoU), and HD95, which are defined as follows:}

\begin{equation}
\label{eq:1}
\text{F1-score} = 100 \cdot \frac{2 \cdot TP}{2 \cdot TP + FP + FN}
\end{equation}

\begin{equation}\label{eq:2}
\text{IoU} = 100 \cdot \frac{TP} {TP + FP + FN}
\end{equation}

\textcolor{black}{where $TP$ is True Positives, $TN$ is True Negatives, $FP$ is False Positives, and $FN$ is False Negatives, all associated with the segmentation classes of the test images.}

\textcolor{black}{The Dice score is a macro metric, which is calculated for $N$ testing images as follow: }
\begin{equation}\label{eq:6}
\text{Dice score} = 100 \cdot \frac{1}{N} \sum_{i=1}^{N} \frac{2 \cdot TP_i}{2 \cdot TP_i + FP_i + FN_i}
\end{equation}
where $TP_i$, $TN_i$, $FP_i$ and  $FN_i$ are True Positives, True Negatives, False Positives and  False Negative for the $i$th image, respectively.

Finally, the HD95 metric is the 95th percentile of the set of distances calculated using the Hausdorff Distance. For the ground-truth mask (G) and the corresponding predicted mask (M), HD95 is defined by:

\begin{equation}\label{eq:7}
\text{HD95}(G, M) = \text{percentile}(h(G, M), 95)
\end{equation}

where the Hausdorff Distance $h(G, M)$ is defined by:

\begin{equation}\label{eq:8}
h(G, M) = \max \left( \sup_{a \in G} \inf_{b \in M} d(a, b), \sup_{b \in M} \inf_{a \in G} d(a, b) \right)
\end{equation}

where $d(a, b)$ is the distance metric between elements \(a\) and \(b\) in sets \(G\) and \(M\), respectively. $\sup$ is the supremum (least upper bound) operation, and $\inf$ is the infimum (greatest lower bound) operation.

For COVID-19 tasks, the four metrics are used for evaluation, as all results are obtained from our experiments. On the other hand, we followed the same metrics used in \cite{afnouch2023bm} and \cite{wang2022uctransnet} for BM segmentation and Gland and Nucleus Segmentation, respectively, as the comparison results are obtained from these two papers for these tasks.

\subsection{Bone-Metastasis Segmentation}
\label{bmseg}
In line with the study by Afnouch et al. \cite{afnouch2023bm}, we present a comprehensive performance analysis of our approach alongside a comparison to existing methodologies. Notably, the original study did not explore transformer-based models. Our investigation extend into the performance of four transformer architectures, namely SwinUnet \cite{cao2022swin}, MTUnet \cite{wang2022mixed}, MISSFormer \cite{huang2022missformer}, and UCTransNet \cite{wang2022uctransnet}. Table \ref{bmres} summarizes the results of five-fold cross-validation, showcasing F1-score, Dice, and IoU indicators. Our approach demonstrates clear superiority across these metrics, even when compared to EDAUnet++ \cite{afnouch2023bm}, an ensemble of five models. Our approach outperforms EDAUnet++ by margins of 1.11\% for F1-score, 1.44\% for Dice, and 1.66\% for IoU, underlining our approach's enhanced performance and computational efficiency. When contrasted with transformer-based architectures, our method consistently outshines them, with UCTransNet emerging as the strongest contender. However, SwinUnet and MTUnet exhibit relatively weaker performance, raising concerns about the generalization capabilities of certain transformer architectures. Further insights are detailed in Table \ref{f1bm}, which presents fold-wise F1-score results and their averages, reaffirming our approach's consistent superiority.

\begin{table}
\centering
\caption{Comparison with state-of-the-art segmentation
methods on BM-Seg dataset. F1-S, D-S and IoU indicators for 5 folds cross-validation are considered following \cite{afnouch2023bm}.}
\label{bmres}
\resizebox{\columnwidth}{!}{%
\begin{tabular}{|l|c|c|c|}
\hline
Model         & F1-S & D-S     & IoU          \\ \hline
U-Net   \cite{afnouch2023bm}    & 79.46& 72.26 & 65.93  \\ \hline
AttUnet  \cite{afnouch2023bm}  & 79.41& 71.76& 65.86 \\ \hline
Unet++  \cite{afnouch2023bm}  & 79.74& 71.99& 66.31  \\ \hline
AttUnet++  \cite{afnouch2023bm} & 80.28 & 72.36 & 67.06  \\ \hline \hline

SwinUnet \cite{cao2022swin} &  61.09   &  39.17  &  44.01   \\\hline   
MTUnet \cite{wang2022mixed} &  58.59  &  44.30   &  41.45  \\\hline   
MISSFormer \cite{huang2022missformer}& 81.44 & 70.42 &  68.73 \\  \hline
UCTransNet \cite{wang2022uctransnet}& 83.62 &  73.88& 71.85  \\ \hline\hline

Hybrid-AttUnet++ \cite{afnouch2023bm}& 82.27 & 75.70 & 69.89  \\ \hline 
EDAUnet++ \cite{afnouch2023bm}&  83.67 & 77.05 &71.92    \\ \hline \hline

\textbf{D-TrAttUnet (Ours)}  & \textbf{84.78} & \textbf{78.49} & \textbf{73.58}   \\ \hline

\end{tabular}}

\end{table}

\begin{table*}
\centering
\caption{F1-score of 5-fold cross-validation experiments for BM segmentation.}
\label{f1bm}

\resizebox{1.3\columnwidth}{!}{\begin{tabular}{|l|c|c|c|c|c|c|}
\hline
Model  & Fold1 & Fold2 & Fold3 & Fold4 & Fold5 &\textbf{Mean}  \\ \hline
U-Net    \cite{afnouch2023bm}    & 80.41& 77.52 & 79.27 & 80.65& 79.44 & \textbf{79.46} \\ \hline
AttUnet  \cite{afnouch2023bm}   & 79.17& 77.79& 79.04 & 80.63& 80.39& \textbf{79.41} \\ \hline
Unet++ \cite{afnouch2023bm} & 80.12& 80.21& 78.44 & 80.27 & 79.67 & \textbf{79.74}  \\ \hline
AttUnet++ \cite{afnouch2023bm} & 80.45 & 80.35 & 79.61 & 80.31 & 80.67 & \textbf{80.28}  \\ \hline  \hline

SwinUnet \cite{cao2022swin} &  60.69   &   60.24  &  60.79   &  59.88  &  63.86  & \textbf{61.09}    \\\hline   
MTUnet \cite{wang2022mixed} &  58.86  &   56.04  & 59.32    &58.15   &  60.57     &  \textbf{58.59}  \\\hline     
MISSFormer   \cite{huang2022missformer}& 81.91 & 82.17 & 78.46 & 81.16 & 83.51 & \textbf{81.44}  \\ \hline
UCTransNet \cite{wang2022uctransnet} & 84.39 & 83.03 & 82.82 &  83.78&  84.07& \textbf{83.62}  \\ \hline  \hline

EDAUnet++ \cite{afnouch2023bm} &84.19 &83.49 &83 &83.96 &83.69&\textbf{83.67} 
\\ \hline  \hline
\textbf{D-TrAttUnet (Ours)} &   \textbf{84.99}&  \textbf{84.64} & \textbf{84.40}  &  \textbf{85.01} &\textbf{84.86}   & \textbf{84.78}
 \\ \hline
\end{tabular}}

\end{table*}


\subsection{Covid-19 Segmentation}
\label{covseg}
For Covid-19 segmentation, both binary and multi-classes tasks are investigated.

\subsubsection{Binary Segmentation}

In this section, we evaluate the performance of the proposed D-TrAttUnet and compare its performance with U-Net \cite{ronneberger_u-net_2015}, Att-Unet  \cite{oktay_attention_2018}, Unet++ \cite{zhou_unet_2018}, CopleNet \cite{wang_noise-robust_2020}, AnamNet \cite{paluru_anam-net_2021}, and SCOATNet \cite{zhao2021scoat} and four recent Transformer-based architectures (SwinUnet \cite{cao2022swin}, MTUnet \cite{wang2022mixed}, MISSFormer \cite{huang2022missformer}, and UCTransNet \cite{wang2022uctransnet}). It should be noted that each experiment was repeated five times. The results shown represent the average $\pm$ the standard deviation of the five runs. 

Table \ref{tab:dataset9}  depicts the obtained results of binary Covid-19 segmentation on Dataset\_2, where  F1-score, Dice, IoU and HD95 indicators are considered.  The comparison shows the superiority of the proposed approach compared with the Baseline architectures, the state-of-the-art approaches and the Transformer-based architectures.  In more details, our proposed D-TrAttUnet approach outperforms the best comparison method by 13.5\% for F1-score (CopleNet), 7.64\% for Dice-score (SwinUnet), and 14.9 for IoU (CopleNet). By looking at the standard deviation of the five experiments, we notice that U-Net, Unet++, Att-Unet, SCOATNet,  CopleNet, MTUnet and  SwinUnet do not have a stable performance. Only AnamNet and MISSFormer approaches have reasonable standard deviation, however, the performance of AnamNet is the lowest in Dataset\_2.  From the results of Dataset\_2, we find that D-TrAttUnet achieves the best performance with stable behavior during different runs.

From these remarks, we conclude that the proposed Dual-Decoders Transformer-CNN based architecture is able to learn from few data, since both CNN and Transformer based features are extracted and fused in the encoding phase, which provides richer global and local features about the infection. The ability to learn from few data is very crucial especially in pandemic, which is  the case in Covid-19 disease.

\begin{table}
 \caption{Performance Evaluation of the Proposed D-TrAttUnet and Different State-of-the-Art Approaches, Including Baseline CNN Architectures (U-Net \cite{ronneberger_u-net_2015}, Att-Unet \cite{oktay_attention_2018}, Unet++ \cite{zhou_unet_2018}), State-of-the-Art Approaches for Covid-19 Infection Segmentation (CopleNet \cite{wang_noise-robust_2020}, AnamNet \cite{paluru_anam-net_2021}, and SCOATNet \cite{zhao2021scoat}), and Recent Transformer-Based Approaches (SwinUnet \cite{cao2022swin}, MTUnet \cite{wang2022mixed}, MISSFormer \cite{huang2022missformer}, and UCTransNet \cite{wang2022uctransnet}) for Binary Covid-19 Infection Segmentation on Dataset\_2.}
\label{tab:dataset9} 
\centering
\resizebox{\columnwidth}{!}{\begin{tabular}{|p{2cm}|p{1.8cm}|p{1.8cm}|p{1.8cm}|p{1.8cm}|}
\hline

 \textbf{Model}  & \textbf{F1-S} &\textbf{D-S} &   \textbf{IoU}&   \textbf{HD95} \\
\hline

U-Net  & 47.36$\pm$14.54 &22.23$\pm$6.51& 32.24$\pm$12.76 &3.24$\pm$0.11\\\hline

Att-Unet  & 50.61$\pm$12.41 & 23.83$\pm$5.16  &  34.82$\pm$11.45 &2.89$\pm$0.09 \\\hline

UNet++  & 55.20$\pm$12.14& 27.01$\pm$5.75& 39.05$\pm$10.96 &3.12$\pm$0.08 \\\hline \hline


CopleNet  & 60.92$\pm$9.16&  26.09$\pm$4.11 &  44.42$\pm$9.44 &2.45$\pm$0.12\\\hline

AnamNet  &38.87$\pm$3.8& 20.13$\pm$1.66&  27.20$\pm$2.91&  5.87$\pm$0.27 \\\hline

SCOATNet  &45.28$\pm$18.46 & 19.87$\pm$7.52&  31.12$\pm$15.56&  2.13$\pm$0.16\\\hline  \hline 

SwinUnet  & 57.75 $\pm$ 4.35 &  29.22$\pm$3.05  &    41.32$\pm$ 3.98&  1.22$\pm$0.12 \\\hline

MTUnet  &  51.83$\pm$ 9.20  &   26.48$\pm$ 4.85 &   35.53 $\pm$ 8.81&  1.27$\pm$0.08  \\\hline 

MISSFormer  & 58.12 $\pm$ 1.86  &  27.90 $\pm$ 0.14 &   40.98 $\pm$1.85&  0.97$\pm$0.06 \\\hline   

UCTransUnet  &  57.83$\pm$4.08  &  26.40$\pm$2.34  &    40.79$\pm$4.13&  0.95$\pm$0.05  \\\hline \hline

\hline

\bf{D-TrAttUnet}& \bf{74.44$\pm$2.38}& \bf{36.86$\pm$2.63}&  \bf{59.34$\pm$3.01} &  \bf{0.88$\pm$0.03}\\\hline
\end{tabular}}

\end{table}

\subsubsection{Multi-classes Segmentation}
Table \ref{tab:MC} summarizes the obtained results of our proposed D-TrAttUnet architecture and the comparison methods for multi-class Covid-19 segmentation. For the GGO infection type, our approach outperforms the comparison architectures. It's worth noting that many of the comparison architectures achieved similar results, with a slight advantage for the UCTransNet architecture. Our architecture achieves better results than the best comparison architecture (UCTransNet) with improvements of 3.15\% for the F1-score, 4.89\% for the Dice score, and 3.61\% for the IoU.
For the Consolidation infection type, it is evident that the performance of all approaches drops compared to GGO. This is mainly because Consolidation infection type is less frequent, as shown in Table \ref{tab:evalsce}. Additionally, it can be challenging to distinguish between consolidation and non-lung tissue, especially when the infection has a peripheral or posterior distribution, which is often the case in Consolidation infection type \cite{kumar_singh_lunginfseg_2021, Laradji_2021_WACV}.
The results in Table \ref{tab:MC} demonstrate that our approach also excels in consolidation segmentation. Specifically, the proposed D-TrAttUnet architecture outperforms the best comparison architecture, UCTransNet, by 8.73\%, 4.6\%, and 8.08\% for F1-score, Dice score, and IoU, respectively. These results underscore the effectiveness of our approach in addressing imbalanced class distribution and the challenge of limited training data, which accurately reflects the real-world scenario of Covid-19 infection.

\begin{table*}
 \caption{Performance Evaluation of the Proposed D-TrAttUnet and Different State-of-the-Art Approaches, Including Baseline CNN Architectures (U-Net \cite{ronneberger_u-net_2015}, Att-Unet \cite{oktay_attention_2018}, Unet++ \cite{zhou_unet_2018}), State-of-the-Art Approaches for Covid-19 Infection Segmentation (CopleNet \cite{wang_noise-robust_2020}, AnamNet \cite{paluru_anam-net_2021}, and SCOATNet \cite{zhao2021scoat}), and Recent Transformer-Based Approaches (SwinUnet \cite{cao2022swin}, MTUnet \cite{wang2022mixed}, MISSFormer \cite{huang2022missformer}, and UCTransNet \cite{wang2022uctransnet}) for multi-classes Covid-19 Infection segmentation (No-infection, GGO and Consolidation).  }
 \begin{center}
\label{tab:MC}
\centering
\scalebox{0.8}{%
\begin{tabular}{|c|l|c|c|c||c|c|c|c|c|}

\hline
 Ex &{\multirow{2}{*}    \textbf{Architecture}}    & \multicolumn{4}{|c|}{\textbf{GGO}}& \multicolumn{4}{|c|}{\textbf{Consolidation}} \\
\cline{3-10}


& & \textbf{F1-S} &\textbf{D-S} &   \textbf{IoU}&\textbf{HD95}& \textbf{F1-S} &\textbf{D-S} &   \textbf{IoU}&\textbf{HD95}  \\
\hline

1&U-Net  \cite{ronneberger_u-net_2015} & 65.81$\pm$1.26 & 50.13$\pm$1.31 & 49.06$\pm$1.41 &  33.58$\pm$2.33 &
31.35$\pm$12.96 & 15.45$\pm$5.66 & 19.26$\pm$8.76&  37.81$\pm$6.81 \\\hline 

2&Att-Unet \cite{oktay_attention_2018}& 64.81$\pm$1.89 & 50.44$\pm$1.35 & 47.97$\pm$2.06 &  34.51$\pm$3.64& 39.04$\pm$6.81 & 19.26$\pm$3.55 & 24.48$\pm$5.31& 37.18$\pm$5.14 \\\hline
3&Unet++ \cite{zhou_unet_2018}& 65.69$\pm$1.29 & 51.65$\pm$4.12 & 48.92$\pm$14.2 &   28.51$\pm$2.94& 31.31$\pm$6.67 & 16.86$\pm$4.48 & 18.75$\pm$4.73& 39.20$\pm$7.12  \\\hline \hline

4&CopleNet \cite{wang_noise-robust_2020}& 60.44$\pm$1.54 & 46.25$\pm$3.13 & 43.33$\pm$1.61  &  37.15$\pm$7.12
& 29.70$\pm$10.29 & 16.46$\pm$4.76 & 17.90$\pm$7.52 &  40.93$\pm$5.46
\\\hline 
5&AnamNet \cite{paluru_anam-net_2021}& 65.10$\pm$ 3.56& 51.69$\pm$4.81 & 48.36$\pm$3.82&    31.11$\pm$4.95
& 31.97$\pm$6.12 & 18.06$\pm$4.61 & 19.18$\pm$4.36 & 38.46$\pm$6.17\\\hline 
6&SCOATNET  \cite{zhao2021scoat}& 65.77$\pm$3.28 & 50.80$\pm$4.63 & 49.09$\pm$3.56 &     34.16$\pm$3.75
& 43.52$\pm$1.67 & 23.32$\pm$2.07 & 27.83$\pm$1.38&  36.47$\pm$4.44\\\hline \hline

7 & SwinUnet \cite{cao2022swin} &  62.74$\pm$2.63  &  42.46$\pm$2.61  &  45.77$\pm$2.83 &  37.54$\pm$4.91 
&  32.2$\pm$6.68  &  19.77$\pm$3.87  &  19.37$\pm$4.60 & 41.30$\pm$7.17 \\\hline 

8 & MTUnet \cite{wang2022mixed} &  57.83$\pm$2.57  &  42.97$\pm$2.78  &  40.72$\pm$2.52  &   36.13$\pm$7.35
&  26.78$\pm$7.39 &  18.24$\pm$ 4.56 &  15.66$\pm$4.74 & 38.88$\pm$7.01 \\\hline 

9 & MISSFormer \cite{huang2022missformer} & 65.66 $\pm$3.06  &  51.57$\pm$4.01  &  48.95$\pm$3.37 & 24.31$\pm$1.33 
&  47.75$\pm$4.77 &  28.02$\pm$2.72  &  31.50$\pm$4.26 & 35.21$\pm$4.06 \\\hline

10 & UCTransNet \cite{wang2022uctransnet} & 67.46$\pm$2.97 & 53.42$\pm$4.24 & 50.97$\pm$3.39 & 31.65$\pm$3.50
& 49.21$\pm$4.27 & 29.41$\pm$3.48 & 32.74$\pm$3.66 & 37.69$\pm$5.69\\\hline \hline

11&\bf{D-TrAttUnet (Ours)}& \bf{70.61$\pm$1.01} & \bf{58.31$\pm$1.37 }& \bf{54.58$\pm$1.21}&   \bf{21.14$\pm$ 1.65}& \bf{57.94$\pm$2.30}& \bf{34.01$\pm$1.40} & \bf{40.82$\pm$2.28}& \bf{33.31$\pm$1.89} \\\hline 
\end{tabular}}
\end{center}
\end{table*}


\subsection{Gland and Nucleus Segmentation}

We conducted an evaluation of our proposed hybrid encoder, TrAttUnet, on two distinct tasks: gland segmentation using the GlaS dataset and nuclear segmentation using the MoNuSeg dataset. In this assessment, we focused solely on the proposed encoder, omitting organ segmentation, and employed a loss function comprising only the cross-entropy loss ($\mathcal{L}_{{LES-CE}}$) and the Dice loss ($\mathcal{L}_{{LES-Dice}}$) for the lesion branch. Our results, summarized in Table \ref{tab:glmo}, showcase D-TrAttUnet's superior performance when compared to state-of-the-art architectures, as reported in \cite{wang2022uctransnet}. These findings underscore the efficiency and generalization capability of our approach across various medical imaging tasks.

\begin{table*}
 \caption{ \textcolor{black}{Hybrid encoder performance evaluation on Gland Segmentation dataset (GlaS) \cite{siri2017gland} and the MoNuSeg dataset \cite{kumar2019multi}. It should be noted that we followed the same evaluation protocol in \cite{wang2022uctransnet}.  }}
 \begin{center}
\label{tab:glmo}
\centering
\resizebox{1.3\columnwidth}{!}{\begin{tabular}{|c|l|c|c||c|c|}

\hline
 Ex &{\multirow{2}{*}    \textbf{Architecture}}    & \multicolumn{2}{|c|}{\textbf{GlaS}}& \multicolumn{2}{|c|}{\textbf{MoNuSeg}} \\
\cline{3-6}


&  &\textbf{D-S} &   \textbf{IoU} &\textbf{D-S} &   \textbf{IoU}  \\
\hline

1& U-Net& 84.87$\pm$1.1  &  74.47$\pm$1.6  & 77.12$\pm$1.9 & 63.45$\pm$2.1 \\\hline 

2& Unet++ & 88.01$\pm$1.1  &  79.03$\pm$1.4  & 75.14$\pm$1.1 & 64.05$\pm$1.4\\\hline 

3& AttUNet& 88.10$\pm$1  &  79.35$\pm$1.2  & 76.14$\pm$1.2 & 63.47$\pm$1.1 \\\hline 

4& MRUNet& 88.43$\pm$1.1  &  80.14$\pm$1.3  & 77.59$\pm$2.1 & 65.01$\pm$2.1 \\\hline

5& TransUNet& 87.88$\pm$0.8  &  79.99$\pm$0.9  & 77.93$\pm$1.2 & 64.75$\pm$1.2 \\\hline

6& MedT& 86.02$\pm$2.4  &  76.45$\pm$3.5  & 76.77$\pm$1.9 & 64.38$\pm$2.9 \\\hline

7& Swin-Unet & 89.79$\pm$0.7  &  82.01$\pm$ 0.8 & 78.01$\pm$0.8 & 63.79$\pm$ 0.9\\\hline

8& UCTransNet& 90.17$\pm$0.5  &  82.85$\pm$1  & 79.01$\pm$0.7 & 64.90$\pm$0.8 \\\hline


9&\bf{TrAttUnet (Ours)}& \bf{92.14$\pm$0.11} & \bf{85.96$\pm$0.17}& \bf{ 79.21$\pm$0.11 }& \bf{ 65.81$\pm$0.12 } \\\hline 
\end{tabular}}
\end{center}
\end{table*}


\subsection{Ablation Study}

\begin{table*}[ht!]
 \caption{Ablation study of Binary Segmentation scenario. The experimental results of Dataset\_2 are summarized with investigating the effectiveness of the following components: Attention Gate (AG), Dual-Decoder (DD) and Transformer Encoder (TrEc).  }
 \begin{center}
\label{tab:abbinary}
\centering
\begin{tabular}{|l|l|ccc|c|c|c|}

\hline
 \multirow{2}{*}\textbf{Ex} & { \multirow{2}{*}   \textbf{Architecture}}   &\multicolumn{3}{|c|}{\textbf{Ablation}}  & \multicolumn{3}{|c|}{\textbf{Dataset\_2}} \\
\cline{3-8}

& &\textbf{AG} &\textbf{DD} &\textbf{TrEc}       & \textbf{F1-S} &\textbf{D-S} &   \textbf{IoU} \\
\hline

1&U-Net (baseline)& \xmark & \xmark & \xmark & 47.36$\pm$14.54 &22.23$\pm$ 6.51& 32.24$\pm$12.76  \\\hline 

2&AttUnet  (baseline)& \cmark & \xmark & \xmark & 50.61$\pm$12.41 & 23.83$\pm$5.16  &  34.82$\pm$11.45  \\\hline 

3&D-TrUnet&\xmark & \cmark & \cmark  & 70.37$\pm$ 3.98 & 36.46$\pm$2.56 & 54.43$\pm$4.80   \\\hline

4&D-AttUnet&\cmark & \cmark & \xmark   &63.43$\pm$ 6.35 & 30.39$\pm$4.08 & 46.76$\pm$6.81 \\\hline
5&TrAttUnet&\cmark & \xmark & \cmark  & 67.33$\pm$6.72 & 32.52$\pm$3.46 & 51.14$\pm$7.76   \\\hline


6&\bf{D-TrAttUnet}&\cmark & \cmark & \cmark & \bf{74.44 $\pm$2.38}& \bf{36.86$\pm$2.63}&  \bf{59.34$\pm$ 3.01} \\\hline

\end{tabular}
\end{center}
\end{table*}

This section delves into a detailed analysis of the significance of various components within the proposed D-TrAttUnet approach. 
To this end, we have chosen to investigate the ablation study to Covid-19 segmentation tasks, encompassing both binary and multi-class scenarios, in order to validate the importance of each component.
Table \ref{tab:abbinary} serves as a concise summary of the results obtained on Dataset\_2 for binary segmentation. Specifically, we focus on assessing the contributions of the Attention Gate (AG), Dual Decoders (DD), and Transformer Encoder (TrEc).
Upon examining Experiments 1 and 2, it becomes evident that the inclusion of the attention gate enhances the performance of the U-Net architecture when applied to Dataset\_2. Subsequently, Experiment 3 involves the removal of the attention gate from our approach. Comparing the results between Experiment 3 and our proposed D-TrAttUnet architecture reveals the paramount importance of the attention gate in our framework. Notably, the results on Dataset\_2 exhibit improvements of 4.07\% for F1-score, 0.4\% for Dice-score, and 4.91\% for IoU, underscoring its substantial role in enhancing segmentation outcomes.
Furthermore, the incorporation of the transformer component within the encoding phase contributes to more robust feature extraction. These enriched features are subsequently channeled through skip connections to the attention gate. Consequently, the attention gate can make more informed selections from the encoder's features and the upsampled features from the previous decoder layer, further enhancing the segmentation process.

In Experiment 4 (refer to Table \ref{tab:abbinary}), we explore the outcomes when the transformer encoder component is omitted from the architecture. This investigation reveals a noticeable decline in results, highlighting the substantial impact of the transformer encoder. To elaborate further, without the transformer encoder, the performance on Dataset\_2 registers a decrease, with metrics such as F1-score, Dice-score, and IoU deteriorating by 11\%, 6.47\%, and 12.58\%, respectively. This depicts the effectiveness of the proposed hybrid encoder, which combines transformer and convolutional layers in the encoding phase, particularly in scenarios with limited data availability, such as during pandemics.
Furthermore, the comparison between Experiment 5 and 6 (as outlined in Table \ref{tab:abbinary}) shows the importance of utilizing Dual-Decoders. Introducing the second decoder for lung segmentation concurrently with infection segmentation results in noticeable improvements in performance, further highlighting the significance of this architectural choice.

\begin{table*}[ht!]
 \caption{Ablation study of Multi-classes Covid-19 Infection Segmentation scenario. The experimental resultsare summarized with investigating the effectiveness of the following components: Attention Gate (AG), Dual-Decoder (DD) and Transformer Encoder (TrEc).  }
 \begin{center}
\label{tab:abMC}
\centering
\scalebox{0.9}{\begin{tabular}{|l|l|ccc|c|c|c|c|c|c|}

\hline
  \multirow{2}{*}\textbf{Ex} & {\multirow{2}{*}    \textbf{Architecture}}   &\multicolumn{3}{|c|}{\textbf{Ablation}}  & \multicolumn{3}{|c|}{\textbf{GGO}}& \multicolumn{3}{|c|}{\textbf{Consolidation}} \\
\cline{3-11}

& &\textbf{AG}&\textbf{DD} &\textbf{TrEc}       & \textbf{F1-S} &\textbf{D-S} &   \textbf{IoU}& \textbf{F1-S} &\textbf{D-S}&   \textbf{IoU}  \\
\hline

1&U-Net  (baseline)& \xmark& \xmark & \xmark & 65.81$\pm$1.26 & 50.13$\pm$1.31 & 49.06$\pm$1.41 & 31.35$\pm$12.96 & 15.45$\pm$5.66 & 19.26$\pm$8.76\\\hline
2&AttUnet  (baseline)& \cmark& \xmark & \xmark & 64.81$\pm$1.89 & 50.44$\pm$1.35 & 47.97$\pm$2.06 & 39.04$\pm$6.81 & 19.26$\pm$3.55 & 24.48$\pm$5.31\\\hline

3&D-TrUnet& \xmark&\cmark & \cmark & 63.77$\pm$1.69 & 49.80$\pm$2.97 & 46.83$\pm$1.82& 50.39$\pm$2.08& 29.19$\pm$4.55 & 33.71$\pm$1.84 \\\hline
4&D-AttUnet& \cmark&\cmark & \xmark & 65.20$\pm$0.95&  51.62$\pm$1.42& 48.38$\pm$1.05 & 51.15$\pm$2.16&29.02$\pm$1.40 &  34.39$\pm$1.98 \\\hline
5&TrAttUnet& \cmark&\xmark & \cmark & 65.69$\pm$1.29&  51.65$\pm$4.12& 48.92$\pm$1.42 & 48.15$\pm$1.75&27.23$\pm$4.52 &  31.73$\pm$1.50 \\\hline

\bf{6}&\bf{D-TrAttUnet}& \cmark&\cmark & \cmark & \bf{70.61 $\pm$1.01} & \bf{58.31 $\pm$1.37}& \bf{54.58$\pm$1.21}& \bf{57.94$\pm$2.31}& \bf{34.01$\pm$1.40} & \bf{40.82$\pm$2.28} \\\hline

\end{tabular}}
\end{center}
\end{table*}

Table \ref{tab:abMC} provides insight into the ablation experiments conducted for multi-class Covid-19 segmentation. Similar to our previous ablation studies in binary segmentation, we examine the significance of three key components: the Attention Gate (AG), Dual Decoders (DD), and Transformer Encoder (TrEc).
Analyzing the results of U-Net and AttUnet (Experiments 1 and 2) for multi-class segmentation reveals the varying impact of the Attention Gate on different classes (Consolidation and GGO). For Consolidation segmentation, the Attention Gate proves highly beneficial, leading to an 8.7\% improvement in F1-score. However, for GGO segmentation, incorporating the Attention Gate results in a slight decrease in performance for metrics like F1-score and IoU.

In contrast, our proposed approach consistently benefits from the Attention Gate for both Consolidation and GGO segmentation (Experiments 3 and 6). From these two experiments, it is noticed that including the AG in our approach enhances the segmentation results for GGO, with improvements of 6.84\% for F1-score, 8.51\% for Dice-score, and 7.75\% for IoU. Similarly, for Consolidation, the Attention Gate leads to improvements of 7.55\%, 4.82\%, and 7.11\% for F1-score, Dice-score, and IoU, respectively. This underscores the critical role of the Attention Gate in identifying crucial features regions from the proposed Hybrid encoder, especially in the complex task of multi-class segmentation.

Examining the fourth and fifth rows of Table \ref{tab:abMC}, it becomes evident that both the Transformer Encoder and the Dual-Decoders are pivotal components in our proposed D-TrAttUnet architecture. Their inclusion results in  performance enhancements, with improvements observed in both  GGO and Consolidation segmentation. Adding Transformer Encoder and the Dual-Decoders leads to substantial improvements in F1-score, particularly notable in the case of Consolidation, where enhancements of about 9.11\% and 12.11\% are achieved, respectively.

Lastly, the final row demonstrates that combining all proposed components in our approach yields the best performance, surpassing the baseline results, particularly for Consolidation. This comprehensive ablation study, encompassing both binary and multi-class Covid-19 segmentation in Tables \ref{tab:abbinary} and \ref{tab:abMC}, underscores the critical importance of each component within our approach.

\section{Qualitative Evaluation and Discussion}
\label{S:6}

\subsection{Visual Comparison}
\label{S:61}
\begin{figure*}[htbp]
\setlength\tabcolsep{1pt}
\settowidth\rotheadsize{Radcliffe Cam}
\begin{tabularx}{\linewidth}{Xp{2pt}Xp{2pt}Xp{2pt}Xp{2pt}Xp{2pt}Xp{2pt}Xp{2pt}X }

       \includegraphics[width = 2.45cm,valign=m]   {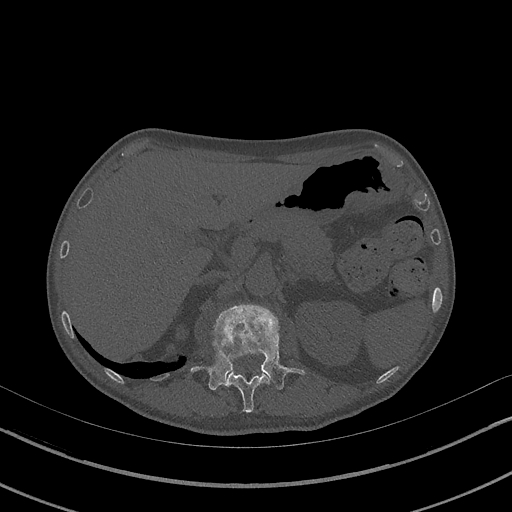} &  
        &  \includegraphics[width = 2.45cm,valign=m] {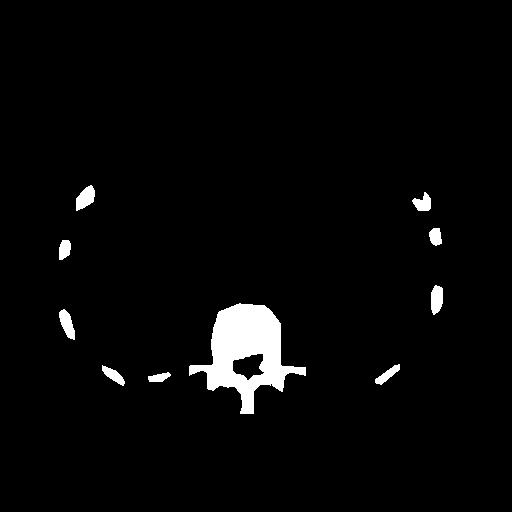}& 
        & \includegraphics[width = 2.45cm,valign=m] {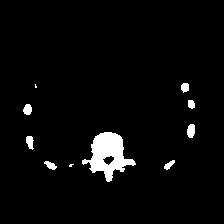}&
        &   \includegraphics[width = 2.45cm,valign=m] {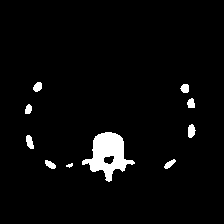}& & \includegraphics[width = 2.45cm,valign=m]{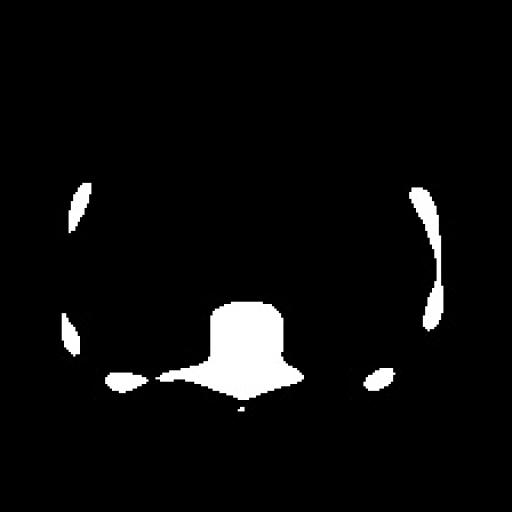}&&
        \includegraphics[width = 2.45cm,valign=m]{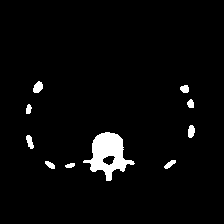}&& \includegraphics[width = 2.45cm,valign=m]{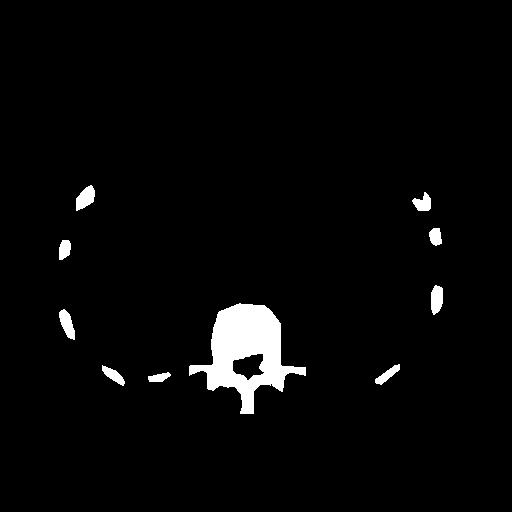} 
        
 \\  \addlinespace[1pt]
 
       \includegraphics[width = 2.45cm,valign=m]   {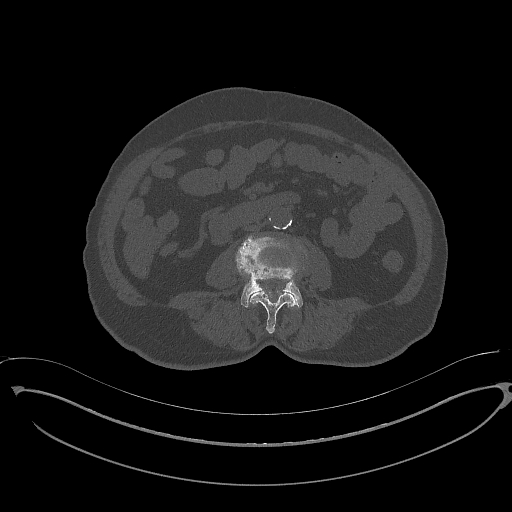} &  
        &  \includegraphics[width = 2.45cm,valign=m] {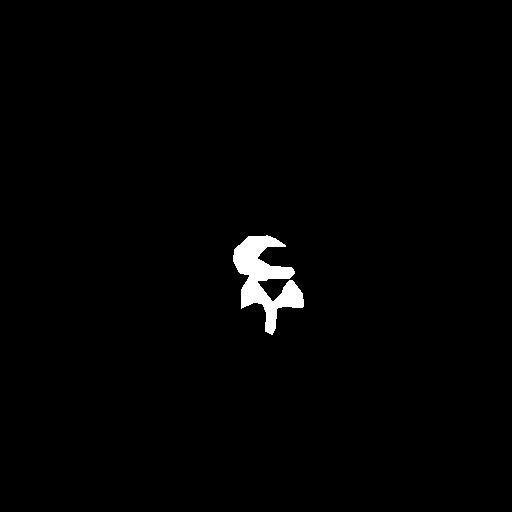}& 
        & \includegraphics[width = 2.45cm,valign=m] {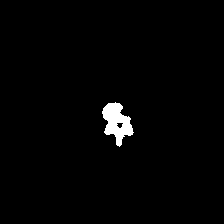}&
        &   \includegraphics[width = 2.45cm,valign=m] {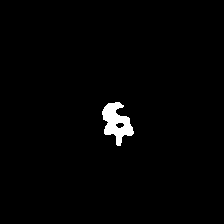}& & \includegraphics[width = 2.45cm,valign=m]{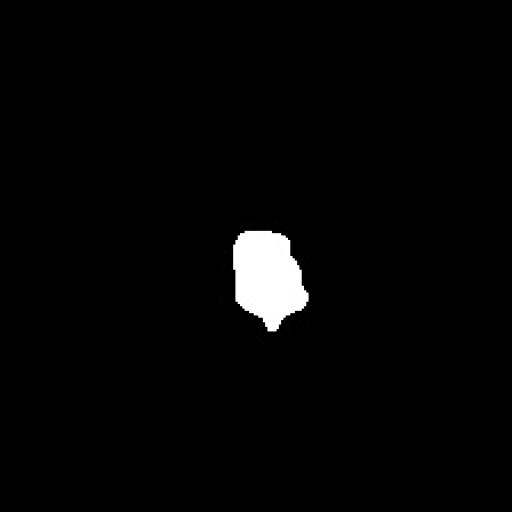}&&
        \includegraphics[width = 2.45cm,valign=m]{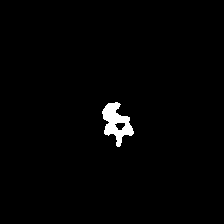}&& \includegraphics[width = 2.45cm,valign=m]{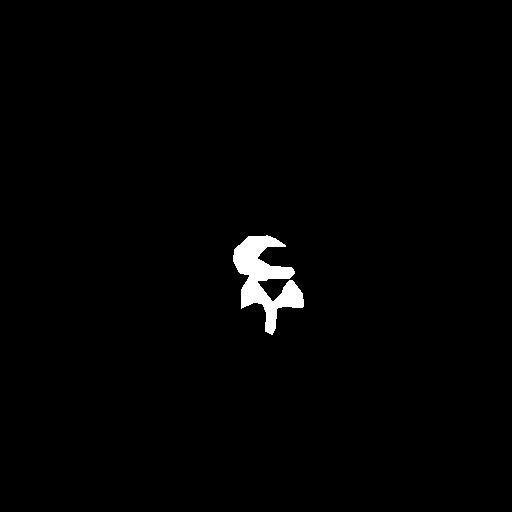} 
        
 \\  \addlinespace[1pt]
 
       \includegraphics[width = 2.45cm,valign=m]   {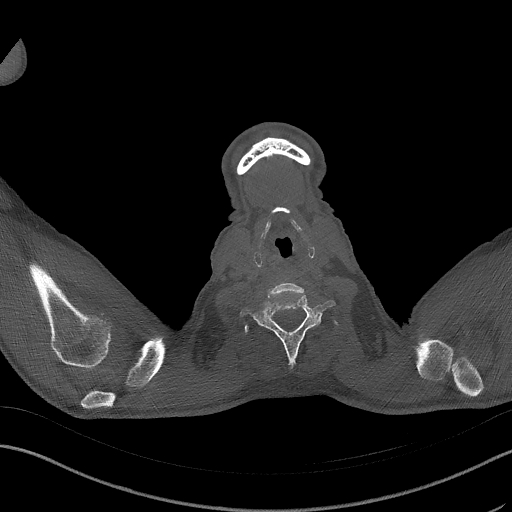} &  
        &  \includegraphics[width = 2.45cm,valign=m] {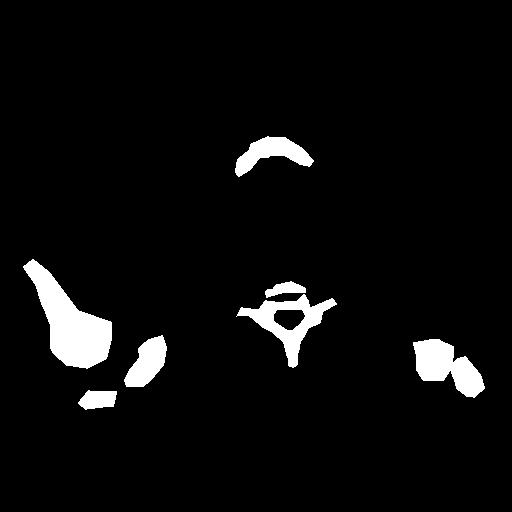}& 
        & \includegraphics[width = 2.45cm,valign=m] {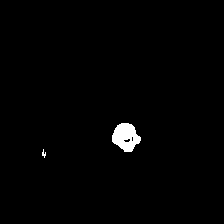}&
        &   \includegraphics[width = 2.45cm,valign=m] {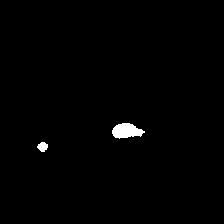}& & \includegraphics[width = 2.45cm,valign=m]{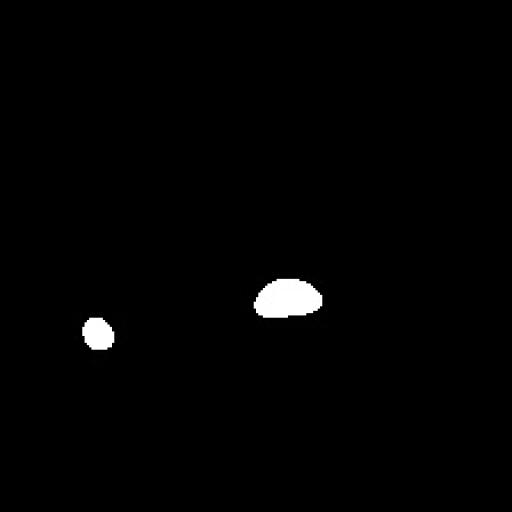}&&
        \includegraphics[width = 2.45cm,valign=m]{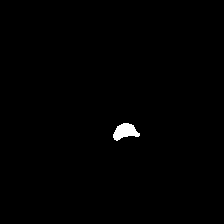}&& \includegraphics[width = 2.45cm,valign=m]{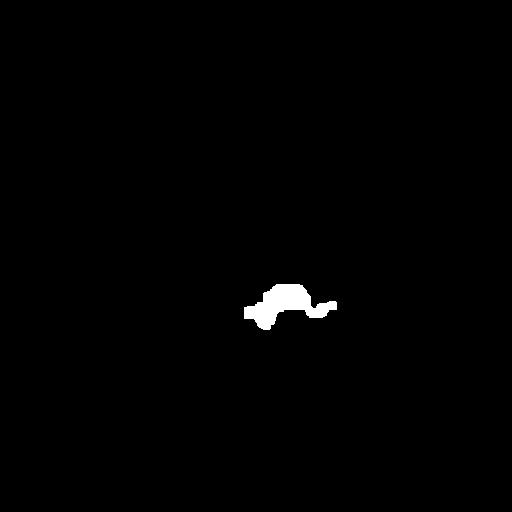} 
 \\  \addlinespace[1pt]
 
       \includegraphics[width = 2.45cm,valign=m]   {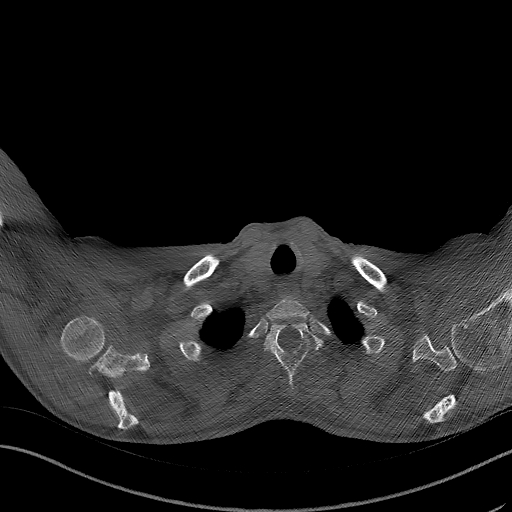} &  
        &  \includegraphics[width = 2.45cm,valign=m] {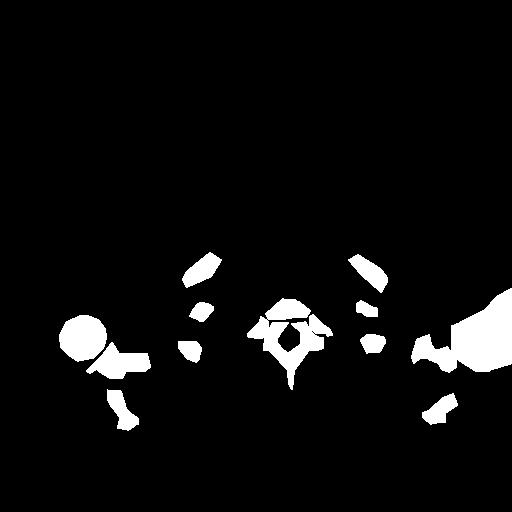}& 
        & \includegraphics[width = 2.45cm,valign=m] {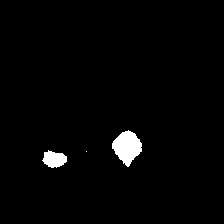}&
        &   \includegraphics[width = 2.45cm,valign=m] {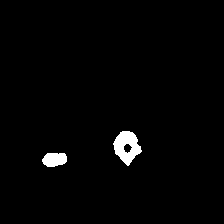}& & \includegraphics[width = 2.45cm,valign=m]{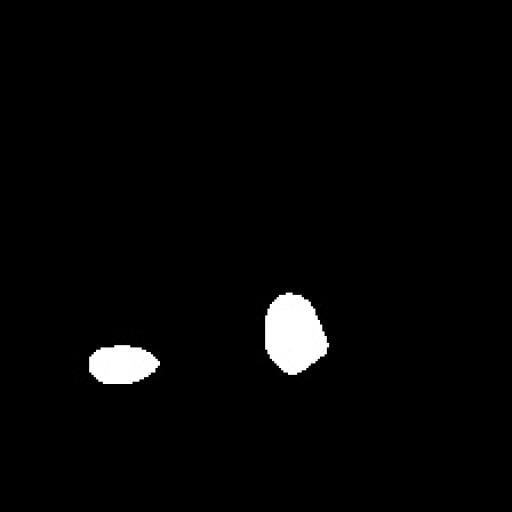}&&
        \includegraphics[width = 2.45cm,valign=m]{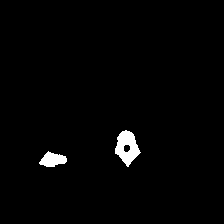}&& \includegraphics[width = 2.45cm,valign=m]{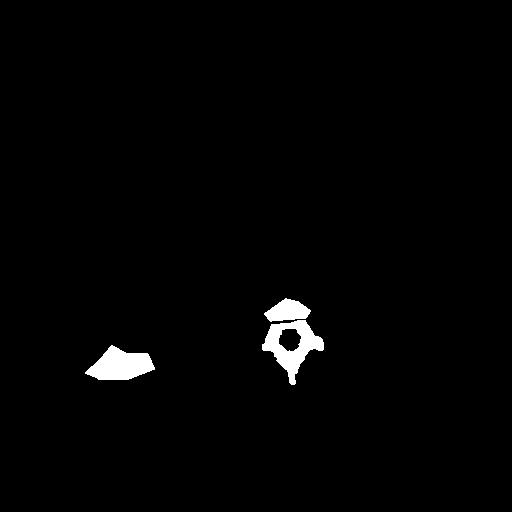} 

 \\  \addlinespace[2pt]
 
       \includegraphics[width = 2.45cm,valign=m]   {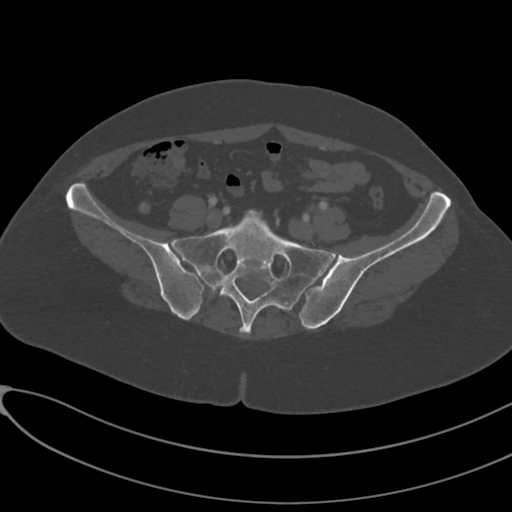} &  
        &  \includegraphics[width = 2.45cm,valign=m] {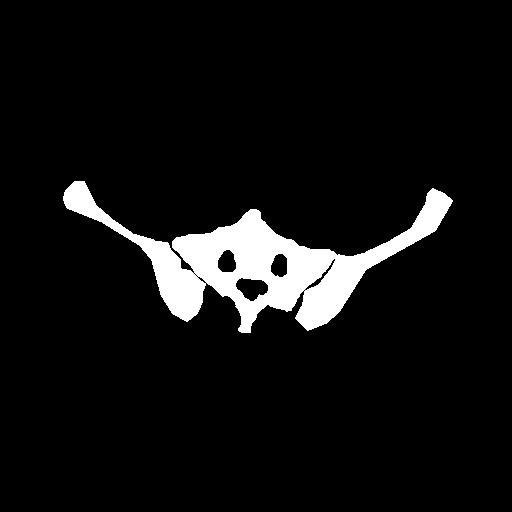}& 
        & \includegraphics[width = 2.45cm,valign=m] {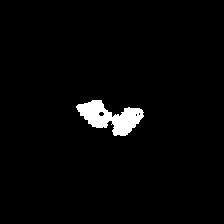}&
        &   \includegraphics[width = 2.45cm,valign=m] {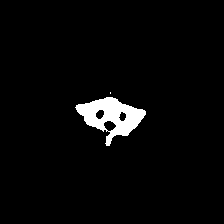}& & \includegraphics[width = 2.45cm,valign=m]{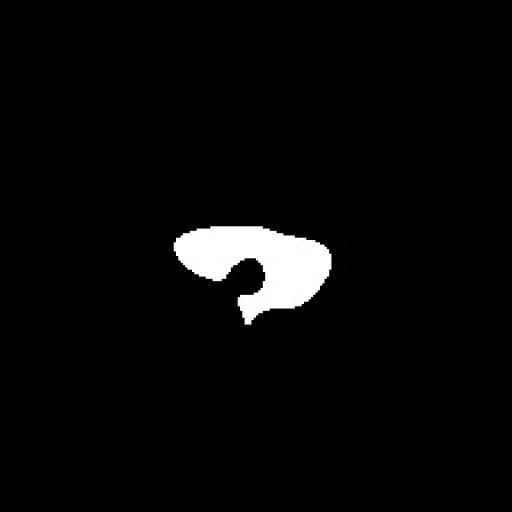}&&
        \includegraphics[width = 2.45cm,valign=m]{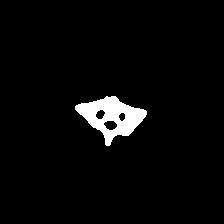}&& \includegraphics[width = 2.45cm,valign=m]{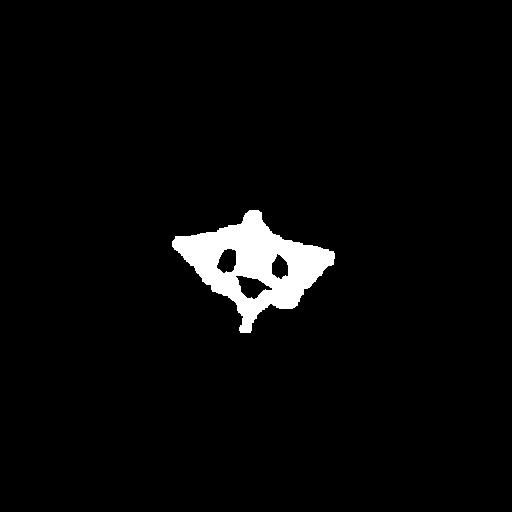}          
         
        \\
        \centering\bf{Slice} &&\centering\bf{Bone Mask}&&\centering \bf{MISSFormer} &&\centering\bf{UCTransNet}&&\centering\bf{EDAUnet+} &&\centering\bf{D-TrAttUnet}&&\centering\bf{GT}
                      
\end{tabularx}
\caption{Visual Comparison of Bone Metastasis Segmentation Models Trained with Different Architectures.  }
\label{fig:compvisbm}
\end{figure*}

\begin{figure*}[htbp]
\setlength\tabcolsep{1pt}
\settowidth\rotheadsize{Radcliffe Cam}
\begin{tabularx}{\linewidth}{Xp{2pt}Xp{2pt}Xp{2pt}Xp{2pt}Xp{2pt}Xp{2pt}Xp{2pt}X }
 
       \includegraphics[width = 2.45cm,valign=m]   {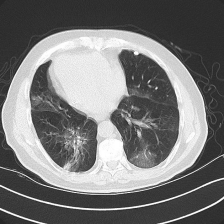} &  
        &  \includegraphics[width = 2.45cm,valign=m] {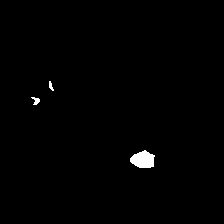}& 
        & \includegraphics[width = 2.45cm,valign=m] {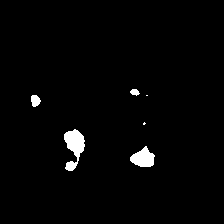}&
        &   \includegraphics[width = 2.45cm,valign=m] {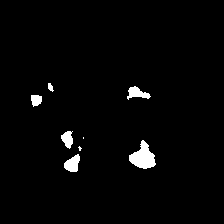}& & \includegraphics[width = 2.45cm,valign=m]{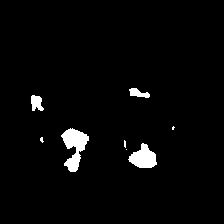}&&
        \includegraphics[width = 2.45cm,valign=m]{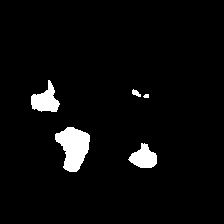}&& \includegraphics[width = 2.45cm,valign=m]{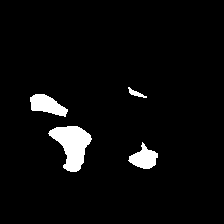}
        
 \\  \addlinespace[1pt]
 
       \includegraphics[width = 2.45cm,valign=m]   {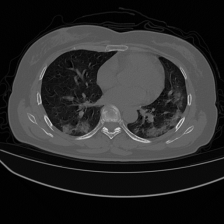} &  
        &  \includegraphics[width = 2.45cm,valign=m] {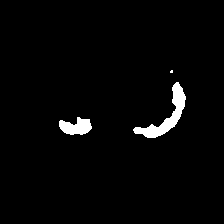}& 
        & \includegraphics[width = 2.45cm,valign=m] {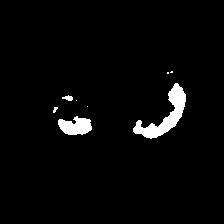}&
        &   \includegraphics[width = 2.45cm,valign=m] {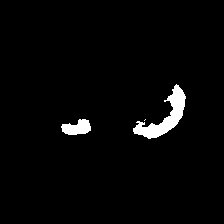}& & \includegraphics[width = 2.45cm,valign=m]{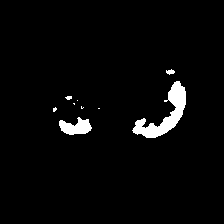}&&
        \includegraphics[width = 2.45cm,valign=m]{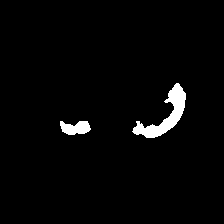}&& \includegraphics[width = 2.45cm,valign=m]{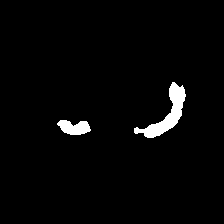}
        
 \\  \addlinespace[1pt]
 
       \includegraphics[width = 2.45cm,valign=m]   {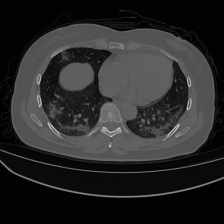} &  
        &  \includegraphics[width = 2.45cm,valign=m] {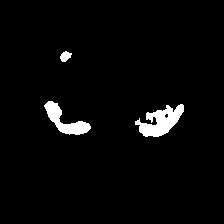}& 
        & \includegraphics[width = 2.45cm,valign=m] {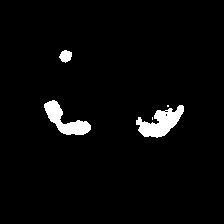}&
        &   \includegraphics[width = 2.45cm,valign=m] {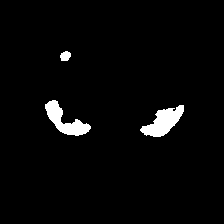}& & \includegraphics[width = 2.45cm,valign=m]{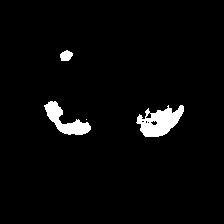}&&
        \includegraphics[width = 2.45cm,valign=m]{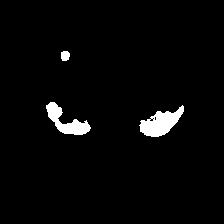}&& \includegraphics[width = 2.45cm,valign=m]{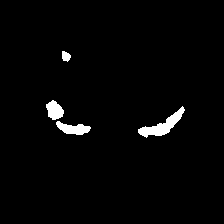}        
 \\  \addlinespace[2pt]
 
       \includegraphics[width = 2.45cm,valign=m]   {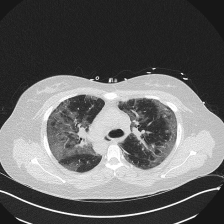} &  
        &  \includegraphics[width = 2.45cm,valign=m] {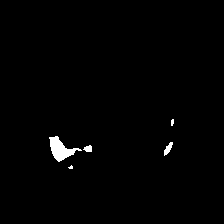}& 
        & \includegraphics[width = 2.45cm,valign=m] {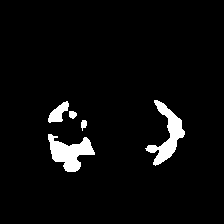}&
        &   \includegraphics[width = 2.45cm,valign=m] {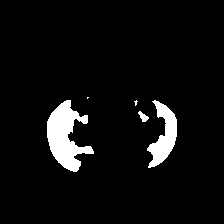}& & \includegraphics[width = 2.45cm,valign=m]{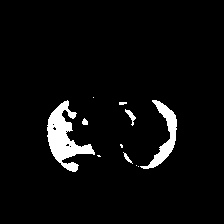}&&
        \includegraphics[width = 2.45cm,valign=m]{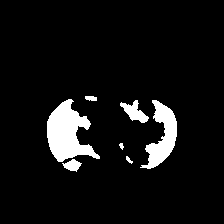}&& \includegraphics[width = 2.45cm,valign=m]{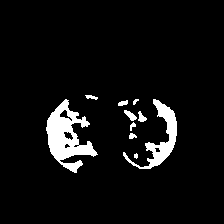}

        \\
        \centering\bf{Slice} &&\centering\bf{Unet++}&&\centering \bf{CopleNet} &&\centering\bf{MISSFormer}&&\centering\bf{UCTransNet}&&\centering\bf{D-TrAttUnet}&&\centering\bf{GT}
                      
\end{tabularx}
\caption{Visual comparison of a segmentation model trained with different segmentation architectures for Binary Covid-19 segmentation using Dataset\_2 and Dataset\_3.  }
\label{fig:compvis}
\end{figure*}

\begin{figure*}[htbp]
\setlength\tabcolsep{1pt}
\settowidth\rotheadsize{Radcliffe Cam}
\begin{tabularx}{\linewidth}{Xp{2pt}Xp{2pt}Xp{2pt}Xp{2pt}Xp{2pt}Xp{2pt}Xp{2pt}X }

       \includegraphics[width = 2.45cm,valign=m]   {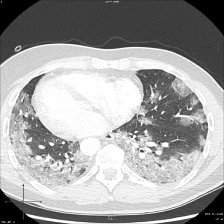} &  
        &  \includegraphics[width = 2.45cm,valign=m] {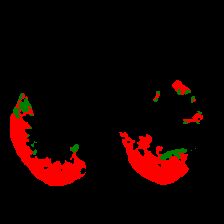}& 
        & \includegraphics[width = 2.45cm,valign=m] {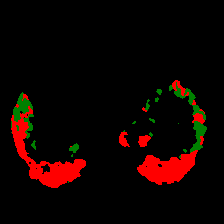}&
        &   \includegraphics[width = 2.45cm,valign=m] {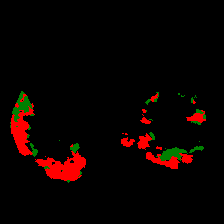}& & \includegraphics[width = 2.45cm,valign=m]{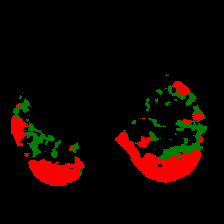}&&
        \includegraphics[width = 2.45cm,valign=m]{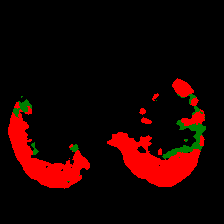}&& \includegraphics[width = 2.45cm,valign=m]{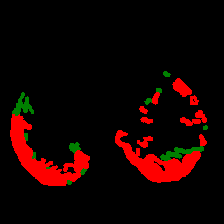}
        
 \\  \addlinespace[2pt]
 
       \includegraphics[width = 2.45cm,valign=m]   {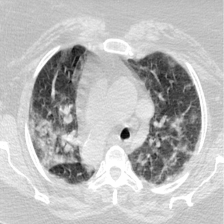} &  
        &  \includegraphics[width = 2.45cm,valign=m] {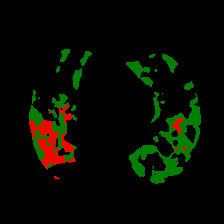}& 
        & \includegraphics[width = 2.45cm,valign=m] {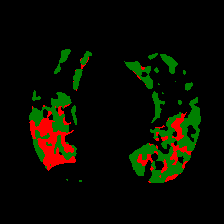}&
        &   \includegraphics[width = 2.45cm,valign=m] {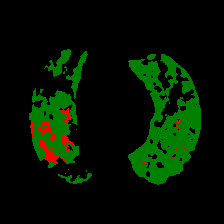}& & \includegraphics[width = 2.45cm,valign=m]{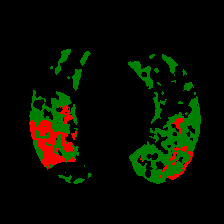}&&
        \includegraphics[width = 2.45cm,valign=m]{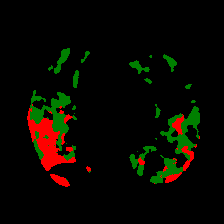}&& \includegraphics[width = 2.45cm,valign=m]{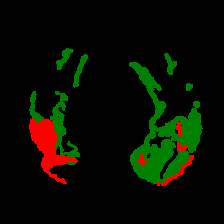}  
        
 \\  \addlinespace[2pt]
 
       \includegraphics[width = 2.45cm,valign=m]   {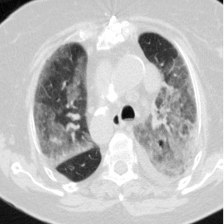} &  
        &  \includegraphics[width = 2.45cm,valign=m] {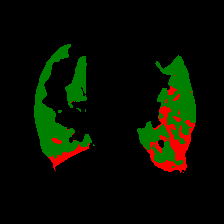}& 
        & \includegraphics[width = 2.45cm,valign=m] {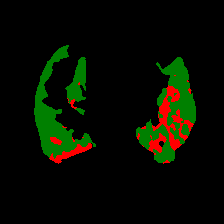}&
        &   \includegraphics[width = 2.45cm,valign=m] {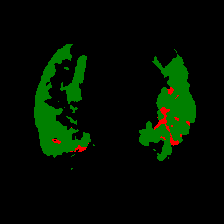}& & \includegraphics[width = 2.45cm,valign=m]{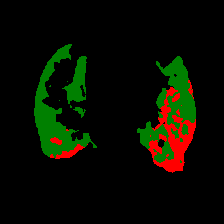}&&
        \includegraphics[width = 2.45cm,valign=m]{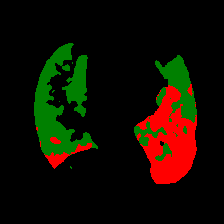}&& \includegraphics[width = 2.5cm,valign=m]{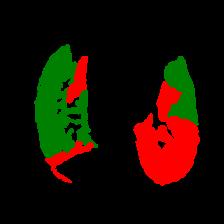}

        \\
        \centering\bf{Slice} &&\centering\bf{AttUnet}&&\centering \bf{SCOATNet} &&\centering\bf{MISSFormer}&&\centering\bf{UCTransNet}&&\centering\bf{D-TrAttUnet}&&\centering\bf{GT}
                      
\end{tabularx}
\caption{Visual comparison of a segmentation model trained with different segmentation architectures for Multi-classes (No-infection, GGO and Consolidation) Covid-19 infection segmentation using Dataset\_2. GGO is presented by the Green color and Consolidation by the red color.  }
\label{fig:compvismc}
\end{figure*}

In our study, we not only compared our approach to state-of-the-art architectures but also provided visualizations of predicted masks for three tasks: BM segmentation, Binary Covid-19 segmentation, and Multi-class Covid-19 segmentation. These visualizations are available in Figures \ref{fig:compvisbm}, \ref{fig:compvis}, and \ref{fig:compvismc}.

For the BM segmentation task, we compared the predicted masks generated by our approach with those from three competitive methods (Figure \ref{fig:compvisbm}), which had proven to be the top performers, as shown in Table \ref{bmres}. These competitors include MISSFormer \cite{huang2022missformer}, UCTransNet \cite{wang2022uctransnet}, and EDAUnet++ \cite{afnouch2023bm}.
In the first two examples, we examined cases where Bone Metastasis had infected all bone regions within the slice. A closer inspection of the predicted masks revealed that most approaches were successful in highlighting the infected regions. However, it was evident that MISSFormer and EDAUnet++ struggled to capture segmentation details accurately. In contrast, our approach and UCTransNet excelled in matching the details present in the ground-truth masks.
The remaining three examples represented slices where only a portion of the bone was infected by BM, which is a particularly challenging aspect of BM segmentation. In the third example, the comparison methods incorrectly segmented a part of the bone as a Bone Metastasis lesion. In contrast, our approach accurately matched the ground-truth mask in these scenarios. The last two examples demonstrated our approach's ability to capture intricate lesion details effectively, showcasing the efficiency of our proposed compound encoder, which integrates Transformer and CNN features.
In summary, the visualizations and comparisons reaffirm the effectiveness of our approach in BM segmentation, particularly in challenging cases, and highlight its ability to capture intricate lesion details efficiently.

On the other hand, the visualized masks of the comparison methods for the binary and multi-classes Covid-19 infection segmentation are: Unet++ \cite{zhou_unet_2018}, CopleNet \cite{wang_noise-robust_2020}, MISSFormer \cite{huang2022missformer} and UCTransUnet \cite{wang2022uctransnet}, and for multi-classes segmentation task are  Att-Unet \cite{oktay_attention_2018}, SCOATNet \cite{zhao2021scoat}, MISSFormer \cite{huang2022missformer} and UCTransUnet \cite{wang2022uctransnet}, which showed a competitive performance with our proposed approach (see Section \ref{covseg}).

The four visualized examples in Figure \ref{fig:compvis} are from the binary segmentation experiments of Dataset\_2.  The first example shows a case in which infection has spread to both lungs and appears as a GGO and small consolidation region at the bottom of the right lung. The comparison between the Unet++ mask and the ground truth (GT) shows that the Unet++ architecture fails in segmenting most of the infection regions. The  CopleNet, MISSFormer and UCTransNet masks show improved segmentation performance compared to Unet++. However,  these architectures still miss some infected regions or segment lung tissues as infection instead. The mask of our proposed approach shows high similarity with  GT in term of the number of regions and their global shape. Both examples 2 and 3 are cases where the infection has a peripheral distribution. The visualized masks show that the proposed D-TrAttUnet is the best architecture consistent with the ground truth. The fourth example depicts a severe case where the infection has spread to most of the lung regions. The visualized masks  exhibit that our proposed architecture  performs better than the comparison architectures. 

Figure \ref{fig:compvismc} consists of the visualization of three examples masks using our approach and the comparison architectures for multi-classes Covid-19 segmentation. The first example shows a mixture case of GGO and Consolidation, where most of the infected regions consist of consolidation and small GGO regions are attached to the consolidation regions. Unlike the masks of the comparison architectures, the mask of our approach has a high similarity to the ground truth mask for both the consolidation and GGO classes. The second and  third examples also represent a case where both GGO and consolidation are present in both lungs. The infected regions with consolidation are mainly in the lower lobes of both lungs and GGO spreads in both lungs with peripheral and posterior distribution. 
The masks of these examples confirm the observation in the first example, as the predicted masks of D-TrAttUnet show a high similarity to the GT masks for both infection types GGO and Consolidation.

Visual exploration has unequivocally demonstrated the robustness and precision of our proposed approach in both BM and COVID-19 segmentation tasks, unequivocally showcasing its efficacy in capturing critical details and surpassing competing state-of-the-art methods.

\subsection{Model Size and Inference Time Comparison}
\label{S:63}

In this section, we investigate the number of parameters, number of FLOPs, and inference times of our approach in comparison with others. Table \ref{tab:numtime} summarizes these comparisons. It is noteworthy that our approach exhibits a similar number of parameters and FLOPs as many state-of-the-art architectures. Compared to baseline architectures such as U-Net and AttUnet, our proposed architecture has a higher parameter count due to the inclusion of the Hybrid Encoder, which has proven its efficiency in handling the complex task with very limited training data.

Moreover, it is important to mention that the second decoder for the Organ segmentation task could be omitted during the testing phase. Since the two decoders operate in parallel and are entirely independent of each other, removing the organ segmentation decoder does not affect the overall functionality. Despite the larger parameter count, our approach still delivers competitive inference times. In fact, the inference time for a batch size of 50 slices is less than half a second, making it suitable for real-time scenarios.

\begin{table}
 \caption{Number of parameters of different architectures and Testing Time for a batch size of 50 slices.} 
\label{tab:numtime}
\centering
\resizebox{\columnwidth}{!}{\begin{tabular}{|p{2.5cm}|p{2cm}|p{2cm}|p{1.5cm}|}
\hline

\textbf{Architecture}& \textbf{Number of FLOPs}& \textbf{Number of Parameters}  & \textbf{Inference Time} \\\hline
 
U-Net& 10.73& 7.85 M  &89ms      \\ \hline
AttUnet& 11.05& 7.98 M   &102ms   \\ \hline 
Unet++& 26.51& 9.16 M  &244ms   \\ \hline 

CopleNet& 12.58& 10.52 M &95ms      \\ \hline
AnamNet& 19.48& 15.63 M  &117ms   \\ \hline 
SCOATNET& 29.75& 40.21 M  & 407ms   \\ \hline  \hline 

SwinUnet& 15.12& 41.38 M  & 160ms     \\ \hline
MTUnet& 44.73& 79.07 M  & 629ms  \\ \hline 
UCTransNet& 32.94&  66.43 M & 423ms   \\ \hline 
MISSFormer& 7.21&  42.46 M  & 223ms   \\ \hline  \hline

D-TrAttUnet& 28.47& 70.13 M  & 475ms     \\ \hline

\end{tabular}}
\end{table}

\section{Conclusion}
\label{S:7}

In this research paper, we introduce a novel approach for medical imaging segmentation tasks, blending the power of Convolutional Neural Networks (CNNs) and Transformers. Our proposed D-TrAttUnet Encoder merges CNN and Transformer layers to extract more comprehensive local, global, and long-range dependency features essential for precise medical imaging segmentation. Notably, many medical lesions target one or multiple body organs. Therefore, our D-TrAttUnet architecture features a Dual-Decoder system, enabling simultaneous segmentation of both the lesions and organ regions. Each decoder includes attention gates, linear upsampling, and convolutional blocks.

To assess the performance of our approach, we tackled a range of challenging medical imaging segmentation tasks. These included Bone Metastasis, Binary and Multi-class COVID-19 infection segmentation, as well as Gland and Nucleus Segmentation. Our proposed D-TrAttUnet architecture consistently outperformed state-of-the-art methods in Bone Metastasis and COVID-19 segmentation tasks. Furthermore, the hybrid encoder we introduced demonstrated remarkable efficiency in Gland and Nucleus Segmentation, surpassing existing state-of-the-art solutions.

Our experimental results underscore the significance of employing attention gates within a compound CNN-Transformer encoder, showcasing the advantages of our approach over a CNN-only encoder. An ablation study emphasized the importance of each component within our proposed approach, while the integration of all elements yielded superior performance, marked by stability and consistent results across both evaluated tasks. The visual explorations provided further evidence of our approach's robustness and precision, clearly highlighting its efficacy in capturing crucial details and surpassing competitive state-of-the-art methods.




\end{document}